\documentclass[aps,pre,twocolumn,superscriptaddress]{revtex4-1}  
\usepackage{graphicx,amsmath,epsfig,psfrag,verbatim,color,enumitem,float}
\usepackage[]{xcolor}  %Blue for new; Red for heavily edited
\usepackage{subfigure}
\usepackage{epstopdf}
\usepackage{comment,datetime}
\usepackage[normalem]{ulem}  % EC-02-JUL-2019 to get strikeout 
\usepackage[export]{adjustbox}
\begin{document}

\title{Correlative mapping of local hysteresis properties in VO$_2$}

\author{Melissa Alzate Banguero}
\affiliation{Laboratoire de Physique et d'\'Etude des Mat\'eriaux, ESPCI Paris, PSL Université, CNRS, Sorbonne Université, 75005 Paris, France}

\author{Sayan Basak}
\affiliation{Department of Physics and Astronomy, Purdue University, West Lafayette, IN 47907, USA}
\affiliation{Purdue Quantum Science and Engineering Institute, West Lafayette, IN 47907, USA}

\author{Nicolas Raymond}
\affiliation{Laboratoire de Physique et d'\'Etude des Mat\'eriaux, ESPCI Paris, PSL Université, CNRS, Sorbonne Université, 75005 Paris, France}

\author{Forrest Simmons}
\affiliation{Department of Physics and Astronomy, Purdue University, West Lafayette, IN 47907, USA}
\affiliation{Purdue Quantum Science and Engineering Institute, West Lafayette, IN 47907, USA}

\author{Pavel Salev}
\affiliation{Department of Physics and Astronomy, University of Denver, Denver, Colorado 80208, USA}
\affiliation{Department of Physics and Center for Advanced Nanoscience, University of California San Diego, La Jolla, California 92093, USA}

\author{Ivan K. Schuller}
\affiliation{Department of Physics and Center for Advanced Nanoscience, University of California San Diego, La Jolla, California 92093, USA}

\author{Lionel Aigouy}
\email{lionel.aigouy@espci.fr}
\affiliation{Laboratoire de Physique et d'\'Etude des Mat\'eriaux, ESPCI Paris, PSL Université, CNRS, Sorbonne Université, 75005 Paris, France}

\author{Erica W. Carlson}
\email{ewcarlson@purdue.edu}
\affiliation{Department of Physics and Astronomy, Purdue University, West Lafayette, IN 47907, USA}
\affiliation{Purdue Quantum Science and Engineering Institute, West Lafayette, IN 47907, USA}
\affiliation{Laboratoire de Physique et d'\'Etude des Mat\'eriaux, ESPCI Paris, PSL Université, CNRS, Sorbonne Université, 75005 Paris, France}

\author{Alexandre Zimmers}
\email{azimmers@espci.fr}
\affiliation{Laboratoire de Physique et d'\'Etude des Mat\'eriaux, ESPCI Paris, PSL Université, CNRS, Sorbonne Université, 75005 Paris, France}


\date{\today}

\begin{abstract}

We have developed a new optical microscopy technique able to track micron-sized surface clusters as temperature is varied. Potential candidates for study include phase separated metal-insulator materials, ferroelectrics, and porous structures. Several key techniques (including autofocus, step motor/cross correlation alignments, single-pixel thresholding, pair connectivity correlation length and image convolution) were implemented in order to obtain a time series of thresholded images. Here, we apply this new method to probe the archetypal phase separated insulator-metal transition  in VO$_2$.  A precise time and temperature series of the insulator-metal transition was achieved, allowing 
us to construct for the first time in this material spatial maps of the 
transition temperature T$_c$.  These maps reveal the formation of micron-sized patterns that are reproducible through multiple temperature sweeps within $\sim$0.6°C, although a few isolated patches showed T$_c$ deviations up to ±2°C.
We also derive maps of the local hysteresis widths $\Delta$T$_c$ and local transition widths $\delta$T$_c$.  
The hysteresis width maps show an average width of $\Delta$T$_c$ =4.3°C, consistent with macroscopic transport measurements, with, however, small regions as low as $\Delta$T$_c$$\sim$[0°C-1°C], and as high as 8°C.  The transition width $\delta$T$_c$ maps shows an average of 2.8°C and vary greatly (from 0°C to 8°C), confirming the strong inhomogeneities of T$_c$ in the subpixel structure. A positive correlation between T$_c$ value and hysteresis width $\Delta$T$_c$ is observed by comparing the  spatial distributions of each map. Finally, individual pixels with unique transition characteristics are identified and put forward. This unprecedented knowledge of the local properties of each spot along with the behavior of the entire network paves the way to novel electronics applications enabled by, {\em e.g.}, addressing specific regions with desired memory and/or switching characteristics,
as well as detailed explorations of open questions in the theory of hysteresis.

\end{abstract}

\maketitle

%\today  
%\\
%\currenttime  ~~(GMT)  (= EST+4 = CET-2) 

\section{Introduction} 

Electronic phase separation commonly emerges in a wide variety of quantum materials such as high-T$_c$ superconductors \cite{McElroy2005}, colossal magnetoresistance manganites \cite{Fath1999}, insulator-metal transition (IMT) materials {\cite{Post2018}}, multilayer rhombohedral graphene {\cite{Shi2020}},{\em etc}. An archetypal example of a phase-separated material is vanadium dioxide, VO$_2$, which undergoes a 1$^{st}$ order IMT at T$_c$ $\sim$68°C {\cite{Morin1959}} ({\em i.e.}, just above room temperature) accompanied by an abrupt several-order-of-magnitude resistivity decrease and monoclinic-to-tetragonal structural change. The exact nature of the transition, whether it is a Peierls transition driven by electron-phonon interactions or a Mott-Hubbard transition driven by electron-electron interactions, is still under debate {\cite{Tomczak2009a}}. In the vicinity of the transition, VO$_2$ exhibits a spatial coexistence of metal and insulator domains that form intricate patterns {\cite{Qazilbash2007}}. Analyzing the shape, characteristic size and scaling properties of those patterns can yield valuable information about the fundamental interactions that drive the transition {\cite{shuo-prl}}. Therefore, understanding and controlling the phase-separate state in quantum materials has become a major research field in recent years {\cite{Coll2019}}.
 
Currently, phase separation imaging in quantum materials reported in the literature mostly comes from scanning probe techniques such as STM {\cite{Fath1999,McElroy2005}} and s-SNIM {\cite{Qazilbash2007,shuo-prl}}. While these methods have a very high spatial resolution, fine temporal resolution remains hard to implement since scanning probes are very time-consuming. Moreover, STM lacks resolution at room temperature and loses registry as the temperature is changed \cite{Gomes2007}. To solve this we have developed a new microscopy method to map out clear and stabilized images of the IMT. This optical method allows the precise filming of the transition with hundreds or even thousands of images taken in quick succession ($\sim$$10$ seconds per final image). This allows us to not only follow fine details in the time evolution of the metal-insulating patches but also to filter out thermal noise if needed.  
We first describe the sample preparation and optical response. We then describe the experimental steps necessary to achieve this mapping. While most steps are straightforward, four new crucial steps were keys to this study: ``Height z focusing'', ``Single pixel time traces'', ``Pair connectivity correlation length'' and ``Time domain convolution''. 
These technical developments allowed us to acquire accurate spatial maps of transition temperature distribution, from which the phase separation patterns can be easily obtained at any given temperature. The T$_c$ maps reveal multiple interesting features including the presence of spots with an extremely large or nearly absent hysteresis of the IMT, a positive correlation between the T$_c$ value and the hysteresis width, and high cycle-to-cycle reproducibility of the transition. The detailed knowledge of local properties is the necessary ingredient to develop and test basic phase separation and hysteresis theories, as well as to gain microscopic understanding of the device performance for practical applications of 
quantum materials.

\section{Methods}
\label{sxn:processingSI}

 \begin{figure*}[ht!]
\centering
 \includegraphics[width=0.73\textwidth]{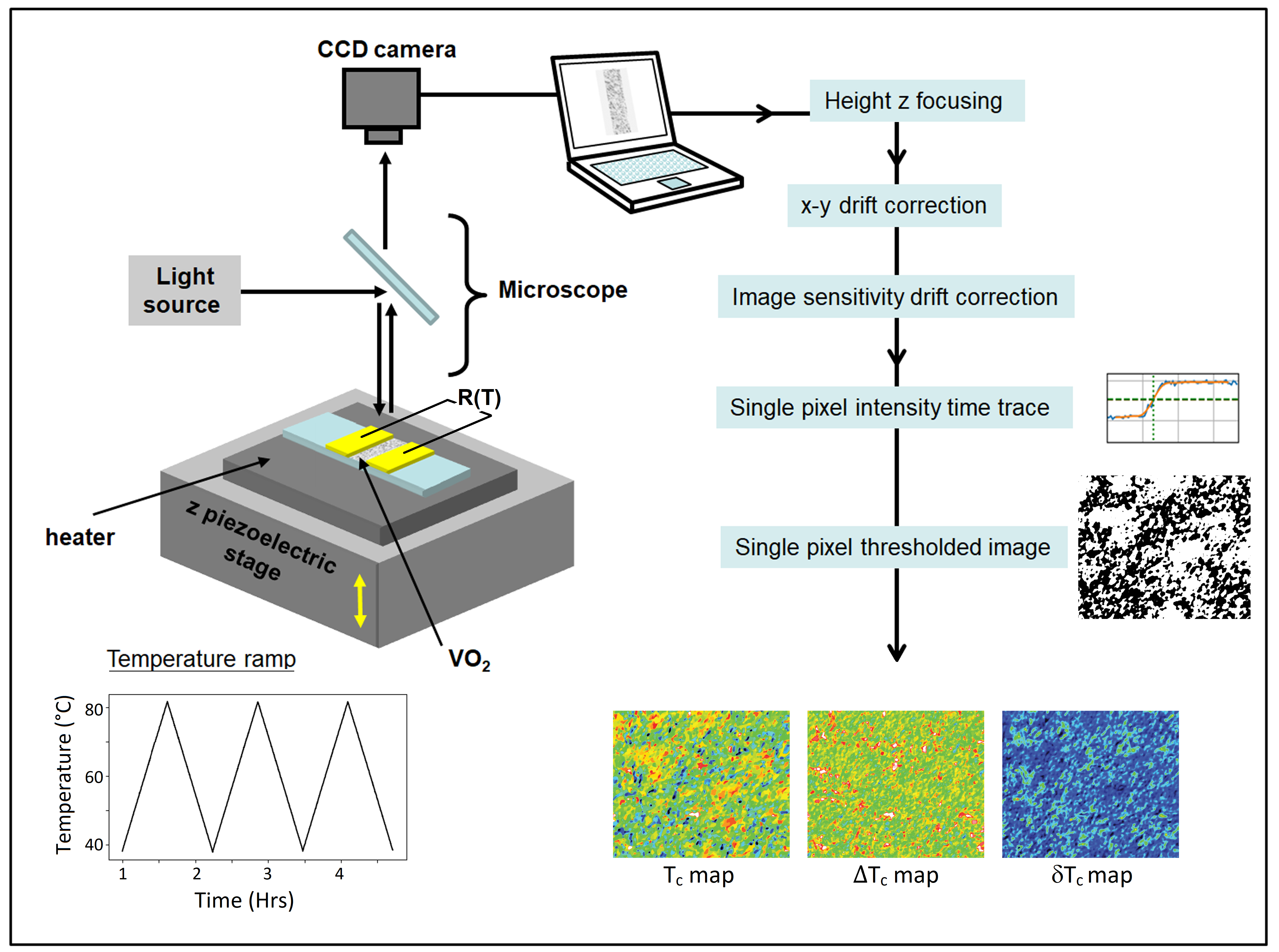}
  \caption{Schematics of the microscope and image analysis 
  created specifically to measure spatial maps of clusters in VO$_2$ during the IMT while recording resistivity R(T) simultaneously.  
  The sample was positioned on a Peltier heater or Linkam Thms350V temperature controller to apply temperature ramps (bottom left). The sample height was varied by steps of 80$nm$ via a piezoelectric actuator placed under it. The best-focused images were chosen post-experiment using an image compression method and Tenengrad function (described in Sec.~\ref{sxn:focusSI}). The height focus of the sample was thus controlled within 80$nm$ throughout the experiment. Fine $xy$ plane drift correction within a single pixel was performed post-experiment (described in Sec.\ref{sxn:focusSI}). Camera sensitivity was normalized throughout the recording (described in Sec.S3 of the SI). Using this fully stabilized image series, black and white thresholds were applied for each pixel individually, accurately determining if it is in the metallic or insulating state (described in Sec.~\ref{sxn:thresholdSI}). We use this information to construct spatial maps of the local transition temperature T$_c$, hysteresis width $\Delta$T$_c$ and transition width $\delta$T$_c$.}
\label{fig:ExperimentalSetup}
\end{figure*}

\subsection{VO$_2$ thin film epitaxy, resistivity, and reflectivity}
\label{sxn:epitaxySI}

Vanadium dioxide thin films were prepared by reactive RF magnetron sputtering of a V$_2$O$_3$ target (\textgreater99.7\%, ACI Alloys, Inc.) on an r-cut sapphire substrate. 
Sample A is 130$nm$ thick and sample B is 300$nm$ thick.  
A mixture of ultrahigh purity (UHP) argon and UHP oxygen was used for sputtering. The total pressure during deposition was 4mTorr, and the oxygen partial pressure was optimized to 0.1mTorr (2.5\% of the total pressure). The substrate temperature during deposition was 600$^o$C while the RF magnetron power was kept at 100W. Grain size in these films is typically found to be 40-130$nm$ in 100-150$nm$ films {\cite{Ramirez2009}}. Grain size is expected to typically be slightly larger in the 300$nm$ film. The
sample is found to have a relative 27\% optical change in the visible range when passing the IMT (see SI Sec.S1 for details). Gold electrodes were deposited on top of the film, separated by 10$\mu m$ (sample A) and 30$\mu m$ (sample B). Both samples showed a clear IMT (see Fig.~\ref{fig:resistance}) above 68$^o$C as evidenced by a drop in resistivity of 4 orders of magnitude {\cite{Zimmers2013}}.

\subsection{Image/temperature recording}
\label{sxn:img_temp_recSI}
The optical experimental setup consists of a VO$_2$ thin film sample placed on a Peltier heater or a Linkam Thms350V temperature controller inside a Nikon optical microscope in epi configuration (both the illumination and reflection of light travel through the same objective). Illumination in the visible range was used (halogen lamp, no filters) {\cite{Currie:17}}. Two surface sample images (sample A 10$\mu m \times$50$\mu m$ and sample B 30$\mu m \times$35$\mu m$) were measured around the focal point of 1mm in the visible range using a $\times$150 magnification dry Olympus objective lens with an optical aperture of NA = 0.9. The theoretical lateral resolution is estimated to be $\delta$r= 1.22$\lambda$/(2 NA) = 370$nm$ in the visible range using the Rayleigh criterion {\cite{Microscoperesolution}}.
Temperature was measured using a Pt100 glued next to the sample. Temperature sweeps (35$^o$C$\ll$T$_c$ to 82$^o$C$\gg$T$_c$ and back) spanning the entire IMT were performed multiple times at a rate of 1°C/min, temperature swept linearly, with temperature and images recorded every $\sim$0.17°C.

\subsection{Height z focusing and x-y drift correction}
\label{sxn:focusSI}

\begin{figure*}[ht!]
\centering
  \includegraphics[width=0.48\textwidth]{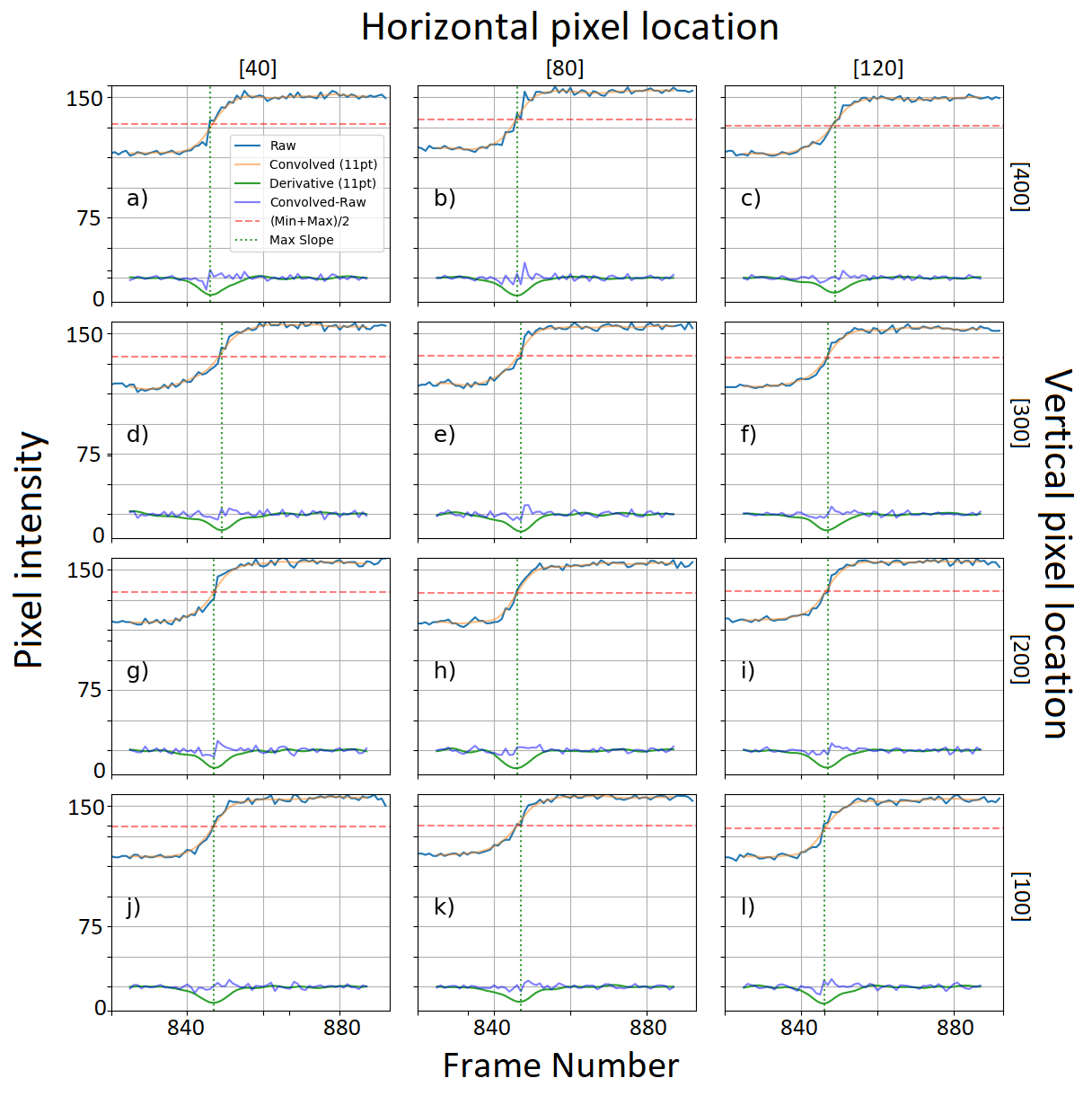}
  \includegraphics[width=0.49\textwidth]{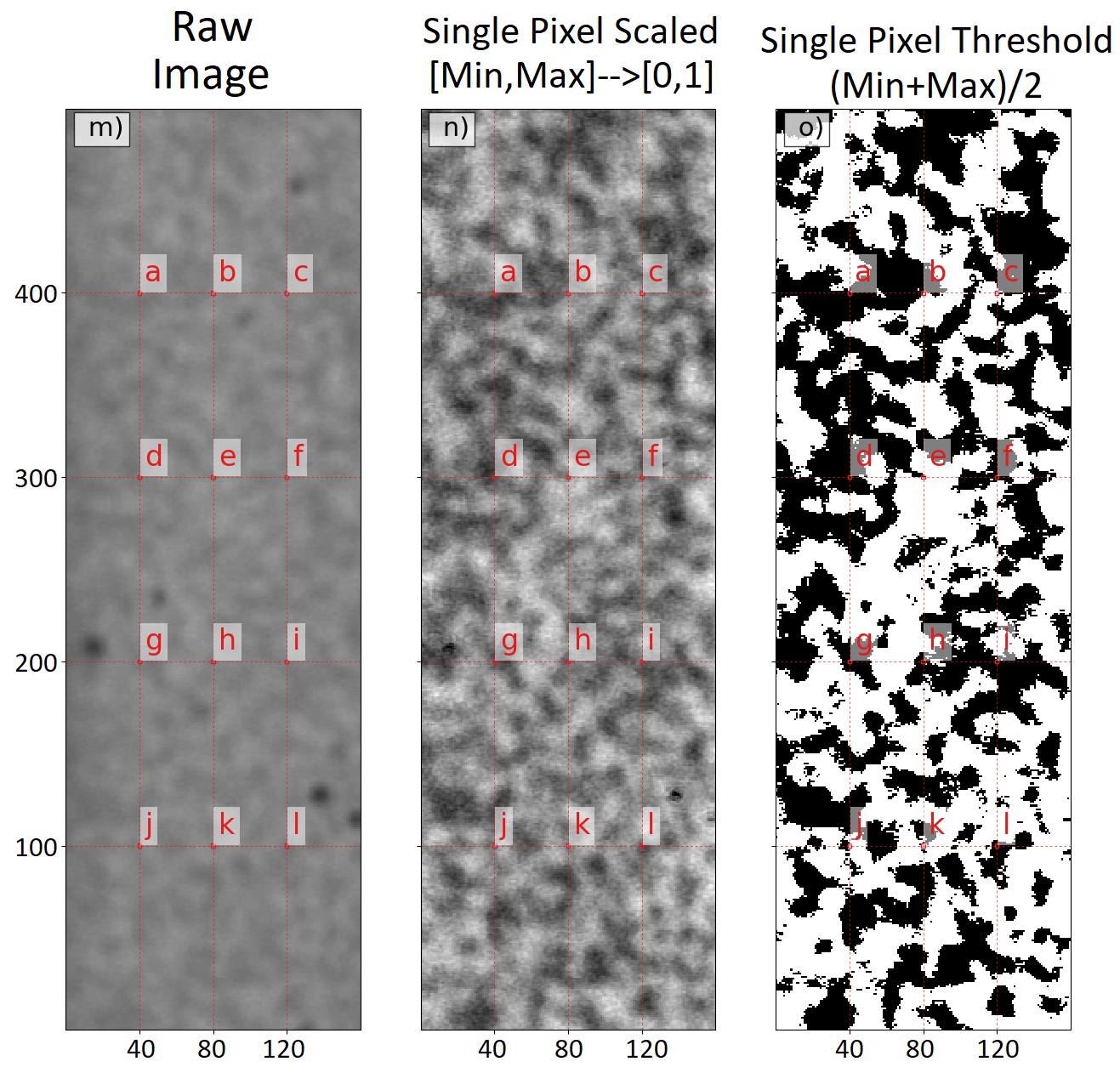}
  \caption{Single pixel intensity normalization and thresholding process.
  (a-l) Representative single-pixel turn-on functions in sample A during cooling. Blue traces are the raw intensity in 8-bit grayscale where 0 is black and 255 is white. The orange traces are smoothed versions of the blue traces, in which we have applied an 11-point Gaussian convolution ($\sigma$=2.5). 
  Purple curves are the difference between the raw (blue) curve and the smoothed version (orange curve). The green curve is a numerical derivative of the blue curve (discussed and used in SI Sec.~S4), taken via a finite difference with a 10-point stencil {\cite{finite-difference}}.
  (m) Raw optical image (frame 847) partway through cooling for VO$_2$ sample A.
  (n) The same image after the intensity is scaled, pixel-by-pixel, such that light pixels are in the insulating phase and dark pixels are in the metallic phase. (o) The same image, with metal and insulator domains, clearly delineated as black and white. Images are 7.3$\mu m$ wide.}
\label{fig:mini_steps}
\end{figure*}

Inevitable temperature dilation of the experimental system during temperature sweeps brings the sample out of focus during temperature sweeps. In order to compensate for this z drift, we employ a ``fuzzy focusing'' technique as follows. 
During the experiment, the sample was continually moved up and down 10$\mu m$  every 10 seconds by a piezoelectric crystal placed under it, in order to bring the sample in and out of focus. A stack of 120 images was recorded this way for each temperature. Over the years, various metrics have been evaluated for selecting the sharpest image in such a stack {\cite{Liu2007,Mir2014,Pertuz2013}}. Some studies focus explicitly on images that don't have sharp contrast {\cite{Liu2016c}}, like the raw images acquired here (see Fig.~\ref{fig:mini_steps}(m)). Most metrics reported perform well in selecting the focused image. We have first chosen one using the compression rate of the recorded images {\cite{Edgett2012}}. This one is based on the intuitive idea that, when very out of focus, the sample surface will look homogeneously gray due to blurring. In this case, the raw recorded Bitmap (BMP) image can be highly compressed in lossless Tiff format using a standard Lempel-Ziv-Welch (LZW) compression protocol {\cite{LZW,LZW2}}, since nearly every pixel is the same. On the contrary, when the sample is in focus, the image contains much more information (since most pixels are different from their neighbors), and the raw BMP image cannot be compressed as much. Using this method, one can determine the most sharply focused image in the stack by selecting the one with the largest Tiff file size {\cite{Note1, PNG}}. 
Among the 62,000 images of sample A acquired during the 14 hour experiment (consisting of 3 major temperature loops and 10 subloops \cite{Basak2022}), we retain the 894 images that are in focus within 80$nm$.

A recent update of the microscope has allowed us to select the best focused image of sample B $during$ the experiment. In the live selection process we have used a computationally faster method based on image gradient using the Tenengrad function \cite{Liu2016c}. Both metrics cited above were vetted using micron-sized gold disks lithographed on a glass substrate where the sharpest image can be defined as the image with the sharpest step function (gold to substrate). 
Using the focusing stack technique, we have also compared the image height on the sample four corners. This allowed us to correct the tilt of the sample (due to sample positioning using thermal paste). The updated setup also uses a piezoelectric PI Pifoc PD72Z1x to move the objective up and down rather than moving the sample placed inside the Linkam stage. The current setup can thus output an image every 10s in focus on the full field of view as a function of temperature.

As the temperature is cycled repeatedly, in addition to drifts along z-axis (perpendicular to the film), there are also drifts in the $xy$ plane (the plane of the film). These thermal drifts were compensated: ($i$) live within  1$\mu m$ using step $xy$ motors below the sample and ($ii$) post experiment using cross correlation to track and realign part of the gold leads which contain imperfections (spots) and rough edges with VO$_2$ (see Fig. \ref{fig:Maps} (a)). Although the lateral image resolution is limited by diffraction and is estimated to be 370nm, the drift compensation tracks each pixel ($\approx$ 37$nm$ wide) on the sample throughout the whole experiment.

The remaining spatial variations we observe in reflected intensity from the VO$_2$ region are primarily due to changes in local reflectivity due to the IMT. However, there can be other contributions to this spatial variation, including effects such as surface height variations from sample warping, variations in film thickness, minor surface defects, and even shadows cast from the 150$nm$ thick gold leads.  There can even be differences in pixel sensitivity in the camera itself.  
Because each of these contributions is independent of temperature ({\em i.e.} constant in time), their effects can be distinguished from that of the temperature driven IMT, as described in the next section. 

\subsection{Single pixel scaled and binary thresholded images}
\label{sxn:thresholdSI}

In order to isolate the changes in local reflectivity which are due to the IMT, we introduce two novel image processing techniques.
We use single pixel time traces to generate {\em single pixel
scaled images} (panel (n) of Fig.~\ref{fig:mini_steps}), as well as {\em binary thresholded images} (panel (o) of Fig.~\ref{fig:mini_steps}, discussed in the following subsections).    
Both types of images begin by considering a full warming or cooling sweep ({\em i.e.} from fully insulating to fully metallic, or {\em vice versa}) to follow the intensity and analyze each pixel individually. As an example, Fig.~\ref{fig:mini_steps} (a-l) shows the raw optical intensity time/frame traces of 12 different pixels during a cooling sweep. See ~\ref{fig:TimeTraceSmall} for the time traces of 1600 pixels from the center of the sample. 
In order to construct a {\em single pixel scaled image}, 
we normalize each individual pixel's 8-bit grayscale intensity time trace with respect to itself, such that its maximum intensity is scaled to 1, and its minimum intensity is scaled to 0.  
The resulting single pixel scaled image is shown in  Fig.~\ref{fig:mini_steps}(n). 
This type of image is a relatively quick way to study
the temperature dependent IMT, as it eliminates temperature-independent spatial variations that are not due to the IMT.

In order to construct a {\em binary thresholded image}
which clearly delineates metal and insulator domains,
we must define a criterion for when each pixel changes from metal to insulator or {\em vice versa}. 
The orange curve in each of the panels (a-l) in Fig.~\ref{fig:mini_steps} is a Gaussian-smoothed version of the raw time trace, using an 11-point Gaussian convolution ($\sigma$=2.5). We use this smoothed time trace of the intensity in order to determine the midway point intensity for each individual pixel (shown by the red horizontal dotted lines). 
We use the pair connectivity correlation length  to justify setting the threshold at midway, as described in the following subsections (Secs.~\ref{sxn:Pair_Connectivity_Correlation_Length} and \ref{sec:threshold}). This allows us to construct {\em binary} black and white images of the metal and insulator domains at each measured temperature, as shown in Fig.~\ref{fig:mini_steps}(o).  
Different pixels go through the midway point at different {\em frame numbers}, and therefore 
at different {\em temperatures}. We use this information to construct spatial maps of the local transition temperature T$_c$ recorded at each pixel revealing the highly spatially-textured nature of the IMT in VO$_2$ {\cite{Qazilbash2007,shuo-prl}}. These
T$_c$ maps, as well as hysteresis width $\Delta$T$_c$ maps and
transition width $\delta$T$_c$ maps, are presented in the experimental results Sec. \ref{sxn:Results}.

\begin{figure}[ht!]
\centering
  \includegraphics[width=1\columnwidth]{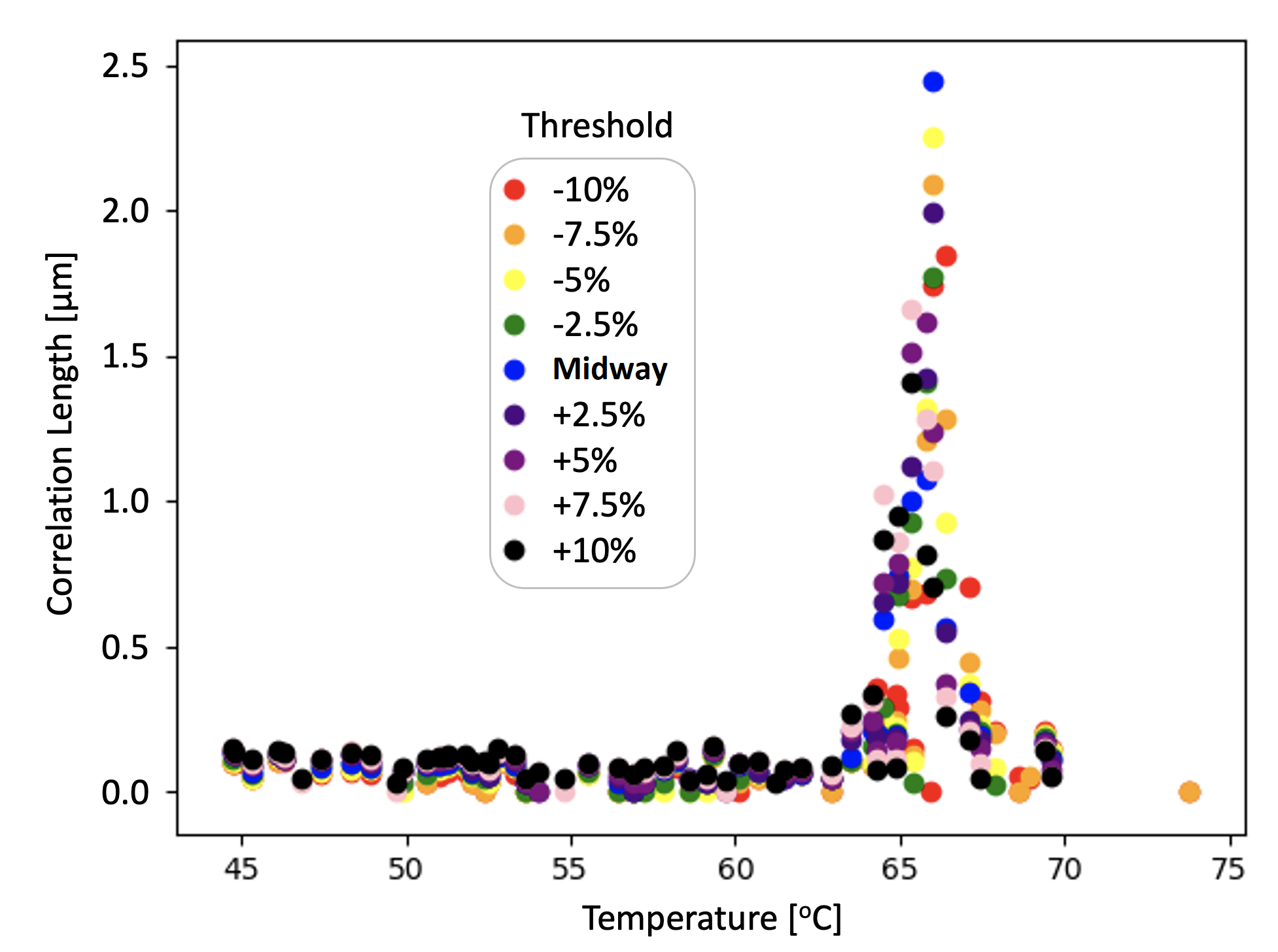}
  \caption{Pair connectivity correlation length $\xi_{\rm pair}$
  {\em vs.} temperature during the warming branch of an extremal hysteresis loop, as a function of different threshold values for determining metal and insulator domains in sample A. The correlation length diverges when the system is closest to criticality.  
\label{fig:xiPair_intensityThreshold}}
\end{figure}

\begin{figure}[h!]
\centering
 \includegraphics[width=0.9\columnwidth]{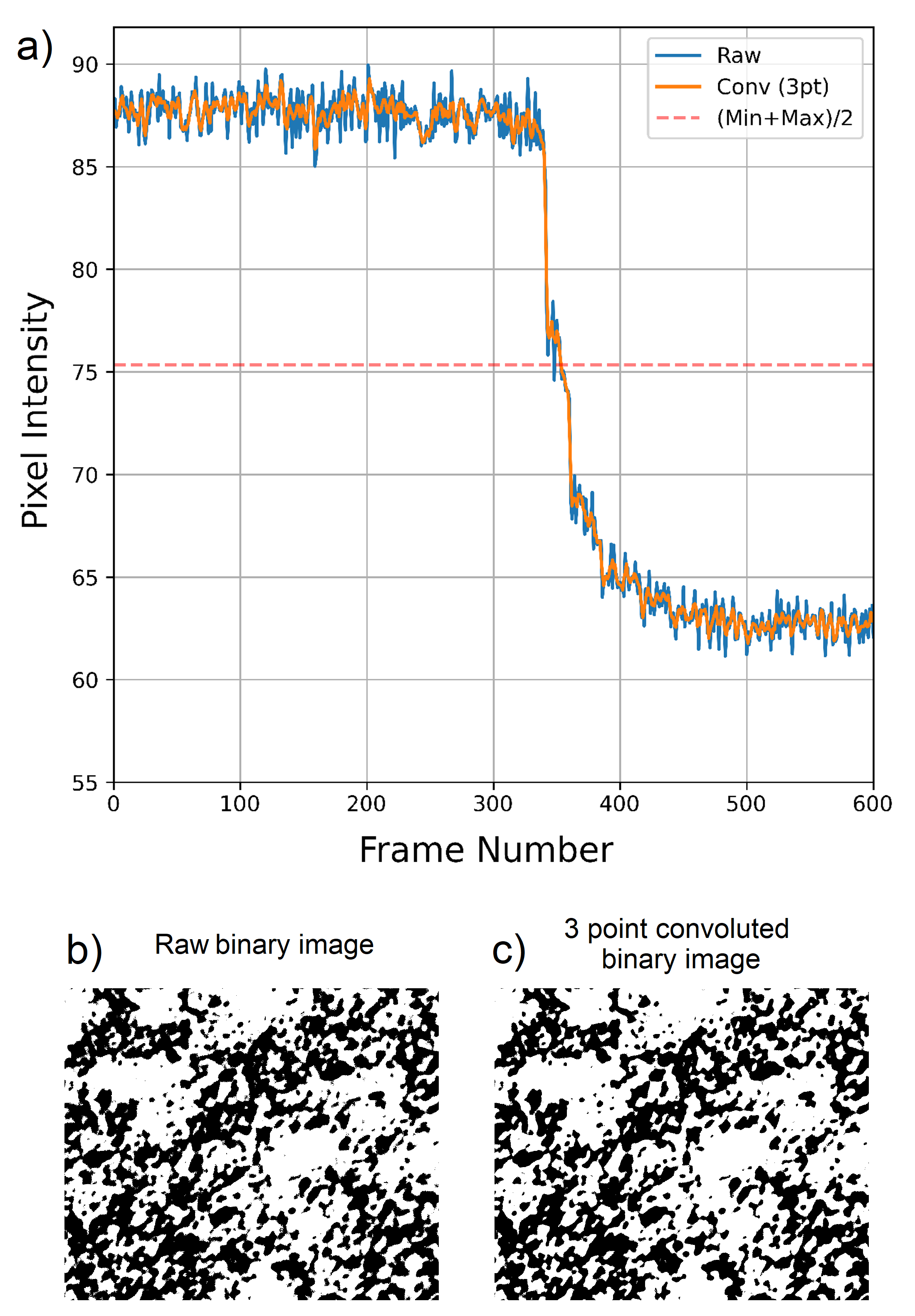}
  \caption{(a) Single pixel time trace of intensity. The blue curve is the raw time trace of the measured optical intensity of pixel (127,734) in sample B. The orange curve is a Gaussian convolution ($\sigma$=2.5) of the same time trace over 3 frames.  The double crossing at the midway is eliminated in the smoothed data set. (b) Binary black and white image (frame 260) of the sample generated by thresholding at midway the single pixel time traces as presented in (a). (c) Smoothed out binary black and white image (frame 260) of the sample generated by thresholding at midway the 3 frame convoluted single pixel time traces  as presented in (a). 
\label{convolution}}
\end{figure}

\subsubsection{Pair Connectivity Correlation Length}
\label{sxn:Pair_Connectivity_Correlation_Length}

As can be seen in the single pixel time traces shown in Fig.~\ref{fig:mini_steps} (see SI Figures.~\ref{fig:TimeTraceSmall} for many more examples), each pixel experiences a definite switch from metal to insulator or {\em vice versa}, consistent with the Ising-type model we have previously developed to describe the IMT in VO$_2$ thin films \cite{shuo-prl,MLpaper}. 
While the Ising model was originally developed to describe magnetic domains of orientation ``up'' or ``down'', here we map ``up'' and ``down'' to metal and insulator domains.  
While the metal-insulator transition is first order, this transition ends in a critical point as a function of quenched disorder.  The influence of that critical point is felt throughout a critical region, which includes part of the first order line in the vicinity of the critical end point.\cite{shuo-prl}
We use the correlation length of the pair connectivity correlation function to determine the threshold between metal and insulator domains.
During the IMT, VO$_2$ metal-insulator
domains form intricate patterns, often becoming fractal due
to proximity to a critical point \cite{shuo-prl}.
At criticality, correlation lengths diverge.  
Away from criticality, the divergence is muted,
although the correlation length still displays a maximum
at the point of closest approach to criticality. 
For example, changing the interaction strength between metal 
and insulator domains to be farther away from criticality, 
or changing the strength of various types of disorder 
farther from criticality causes the correlation length
to go down.  
Similarly, changing the intensity threshold by which
we identify metal and insulator domains also changes this
correlation length.  
In disordered systems, setting an unphysical threshold will not move the system toward criticality, but only away.  Therefore, one way to set the proper threshold
between metal and insulator domains is to maximize the correlation length.

The pair connectivity correlation function is familiar from percolation models, where the corresponding pair connectivity correlation length diverges at the critical point \cite{aharony-book}.  
Coniglio and coworkers showed that the pair connectivity
correlation length also diverges at the critical temperature in the two-dimensional Ising model \cite{Coniglio_1977}.
We have recently shown that the pair connectivity correlation length
also diverges at other Ising critical points, including that of the 
two-dimensional random field Ising model \cite{critical-nematic-hoffman-2021},
as well as on slices of three dimensional models at criticality,
including the clean Ising model \cite{C-3Dx} and the
random field Ising model \cite{critical-nematic-hoffman-2021}.
Near a critical point, the correlation function is power law
at distances less than the correlation length, in this case
$\xi_{\rm pair}$.  
This pair correlation length can be 
calculated directly from an image via \cite{coniglio-cxnlength}:
\begin{equation}
\xi^2_{\rm pair} = \frac{\sum_{i,j} r_{i,j}^2 p^f_{i,j}}{\sum_{i,j} p^f_{i,j}}
\label{eqn:pairCxn}
\end{equation}
where $p^f_{i,j}$ is the likelihood that i and j are in the same finite cluster.  
Another way to view this is as:
\begin{equation}
\xi_{\rm pair} = \sqrt{\left< R_G^2 \right>_f}
\end{equation}
where $R_G$ is the radius of gyration of each connected cluster,
and the average is taken over the finite clusters. 
This quantity diverges at the percolation threshold as:
\begin{equation}
\xi_{\rm pair} \propto \frac{1}{|p-p_c|^{\nu_{\rm pair}}} ~.
\end{equation}
It diverges at clean Ising transitions as:
\begin{equation}
\xi_{\rm pair} \propto \frac{1}{|T-T_c|^{\nu_{\rm pair}}} ~,
\end{equation}
and it diverges at random field Ising transitions as:
\begin{equation}
\xi_{\rm pair} \propto \frac{1}{|R-R_c|^{\nu_{\rm pair}}} ~.
\end{equation}

\subsubsection{Setting Thresholds of Metal and Insulator Signal in Optical Data}
\label{sec:threshold}

In order to know at what intensity to set the threshold 
between metal and insulator in each pixel, we calculate
the pair connectivity correlation length in a series of images,
as a function of different intensity thresholds. 
For this we use the single pixel scaled images as described in the previous subsection. In Fig.~\ref{fig:xiPair_intensityThreshold}, we plot
the evolution of the pair connectivity correlation length (Eqn.~\ref{eqn:pairCxn}) during the warming branch of a hysteresis loop. The blue circles in Fig.~\ref{fig:xiPair_intensityThreshold} have each pixel's threshold set at the midway point
of that particular pixel's intensity. The black circles have each pixel's threshold set higher by an
amount that is $+10\%$ of the difference between the saturated
metal and saturated insulator values of intensity.  The pink
circles have each pixel's threshold set higher by only $+7.5\%$,
and similarly for other colors as denoted in the figure legend. 
Similar to the way the theoretical threshold was set in Ref.~\cite{shuo-prl},
we set the threshold according to the longest correlation lengths.
Since in Fig.~\ref{fig:xiPair_intensityThreshold} the longest correlation length happens for a threshold equal to the average between metal and insulator intensity (the blue circles in Fig.~\ref{fig:xiPair_intensityThreshold}) we use this midway threshold throughout the paper.

\begin{figure*}[ht!]
\centering
 \includegraphics[width=1.0\textwidth]{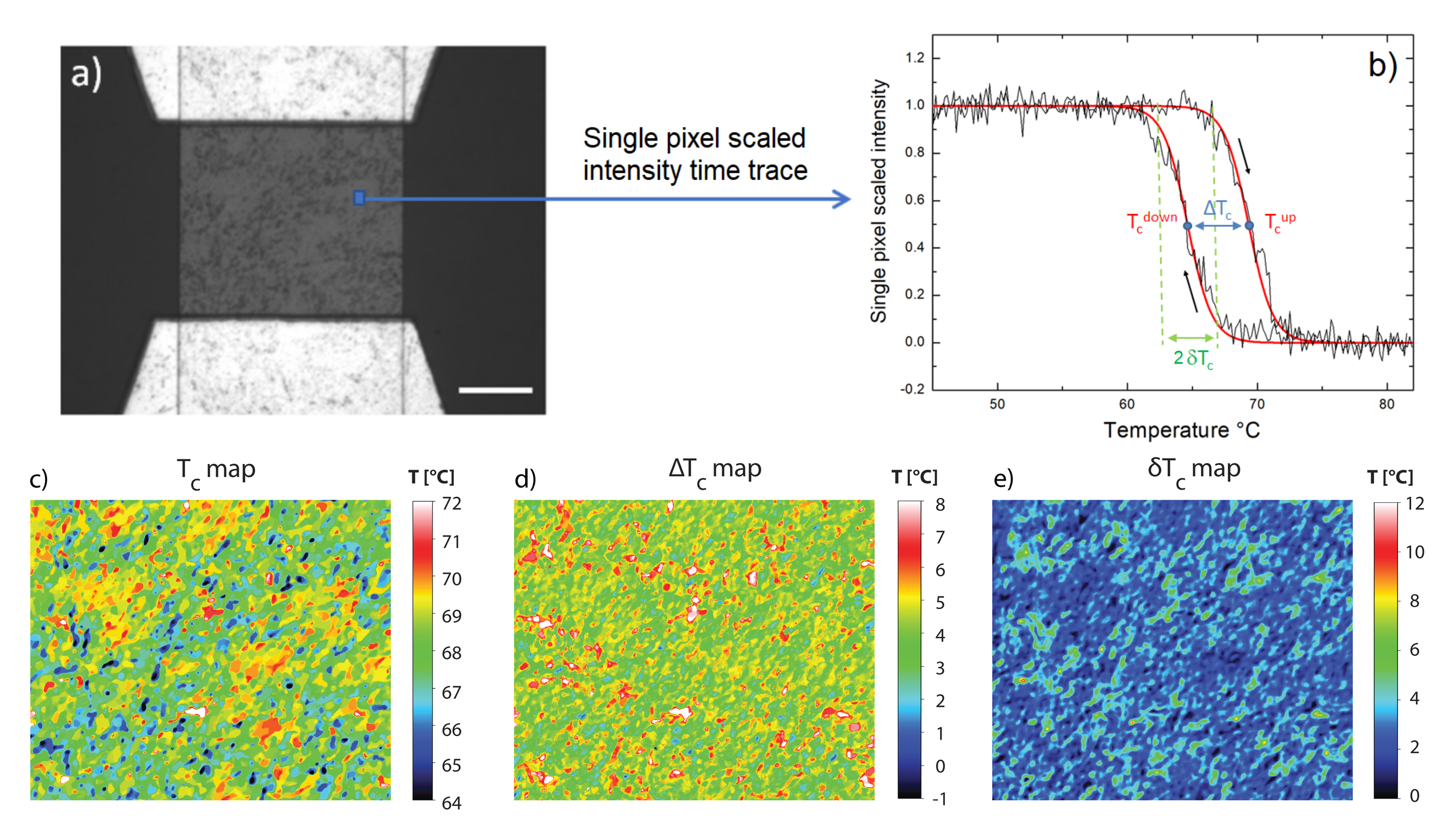}
  \caption{(a) Optical image of VO$_2$ sample B during the insulator (light gray) to metal (dark gray) transition (warming cycle), two gold leads are seen at the top and bottom. These electrodes also display some structure (spots) due to gold surface imperfections. Contrary to VO$_2$ IMT structures seen in this image, gold imperfections do not change with time (see online movie \cite{YouTubeMovie}). Usually these imperfections are purposely washed away using strong image brightness. Here, on the contrary, brightness was set low to see and use these imperfections to autoalign within a pixel the images and thus compensate $xy$ thermal drifts. Sapphire substrate is the dark surface. One can easily see the metal dark patches appearing. Scale bar is 10$\mu m$. (b) Single pixel intensity curve defining critical temperature T$_c$, hysteresis width $\Delta$T$_c$ and transition width $\delta$T$_c$. T$_c$ were determined at midways as explained in the main text. Hysteresis width was determine by taking the temperature differences between heating and cooling cycles T$_c$$^{up}$-T$_c$$^{down}$. Transition width was determined by fitting (smooth curve) the single time trace to a hyperbolic tangent: $-\frac{1}{2}(tanh$($\frac{T-T_c}{\delta T_c}$)-1).  (c) Local critical temperature T$_c$ map, (d) $\Delta$T$_c$ maps, (e) $\delta$T$_c$ map (presented here for the temperature ramping up branch). Image are 27.6$\mu m$ high. Histograms (with mean and standard deviation of maps a), b) and c) are shown in Fig.~\ref{Histograms} }
\label{fig:Maps}
\end{figure*}

\subsection{Time domain convolution}

One of the strong points of obtaining a series of 100-1000 images via this autofocus optical microscope is the possibility of filtering out high frequency noise. A similar technique is used in resistivity experiments that probe samples thousands of times per second. Fig.~\ref{convolution} (a)  compares a raw single pixel time trace to 
a smoothed version in which a 3-point Gaussian convolution ($\sigma$=2.5) has been applied in 
the time domain.  
In this example, the raw single pixel time trace crosses the midway point twice, 
whereas the 3-point convolved curve passes the midway point only once.
Notice that this procedure of filtering high frequency noise in the time domain
greatly suppresses the white noise evident in the spatial domain near the metal-insulator boundaries
derived from the raw time traces (see Fig.~\ref{convolution} (b) and (c) for comparison).  
 This smoothing is useful 
for studying spatial correlations from frame to frame. However, if filtering is not necessary, raw data is used throughout the analysis. This is the case for T$_c$ maps in the section below and ramp reversal memory maps presented elsewhere \cite{Basak2022}.
High frequency noise was filtered in the temperature data taken using the Pt100 by fitting a linear slope through the large temperature sweeps. This matched the internal temperature sensor slope of the Linkam Thms350V temperature controller.

\begin{figure*}[ht!]
\centering
 \includegraphics[width=0.92\textwidth]{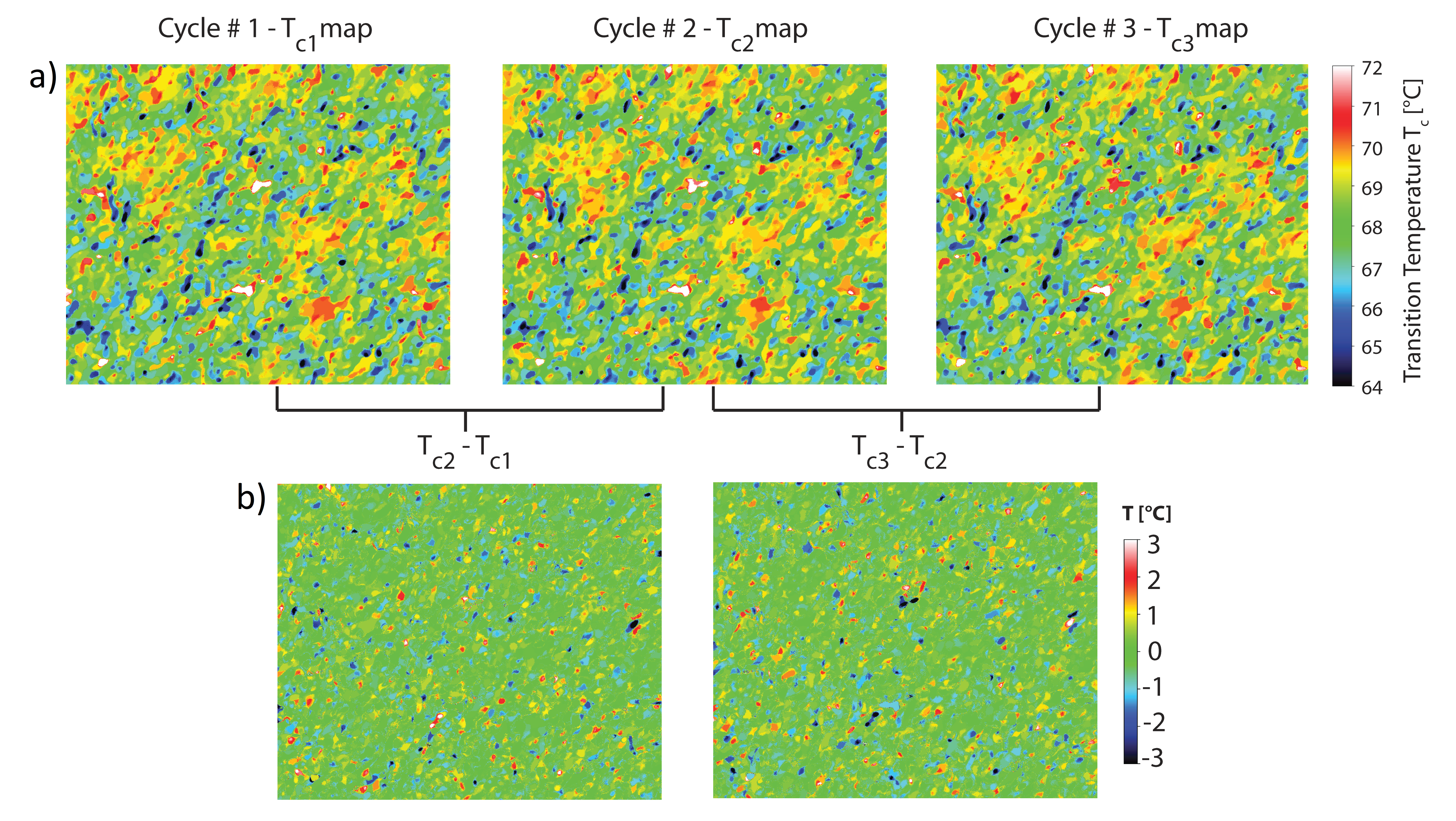}
  \caption{a) Three T$_c$ maps while cycling through the IMT (warming) at 1°C/min. b) Difference maps between cycles. Global patterns are generally reproducible  ($\sigma_{T_c}/T_c = 0.6$°C/68°C$=1\% $).  However some small regions present deviations up to $\pm$2°C. Full histograms (with mean and standard deviation) of maps in b) are shown in Fig.~\ref{Histograms}. Difference map between T$_{c3}$ and T$_{c1}$ (the most separated, time wise, temperature sweeps in this study) and the corresponding histogram are presented in SI Fig.~\ref{Fig:fullTcDiff}. Images are 33.6$\mu m$ x 27.6$\mu m$.}
\label{fig:Repro}
\end{figure*}

\section{Results}
\label{sxn:Results}

\begin{figure*}[ht!]
\centering
\includegraphics[width=1\textwidth, clip]{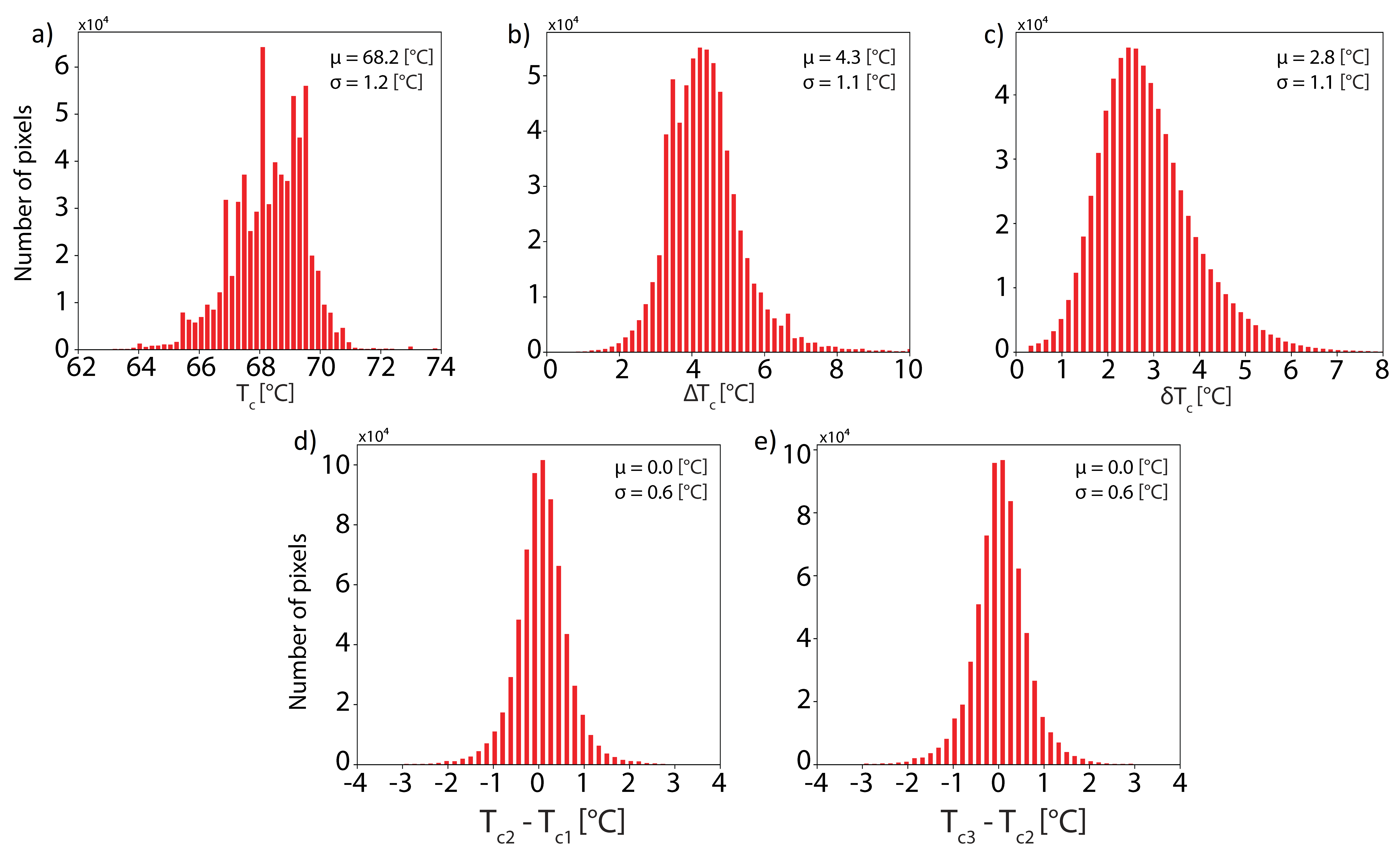}
  \caption{Histograms of maps presented in in Fig.~\ref{fig:Maps} and \ref{fig:Repro}. (a) T$_c$ maps (upon warming); (b) $\Delta$T$_c$ map; (c) $\delta$T$_c$ map and (d) and (e) two difference maps Tc$_2$-Tc$_1$ and Tc$_3$-Tc$_2$}
\label{Histograms}
\end{figure*}

\begin{figure*}[ht!]
\centering
 \includegraphics[width=2\columnwidth]{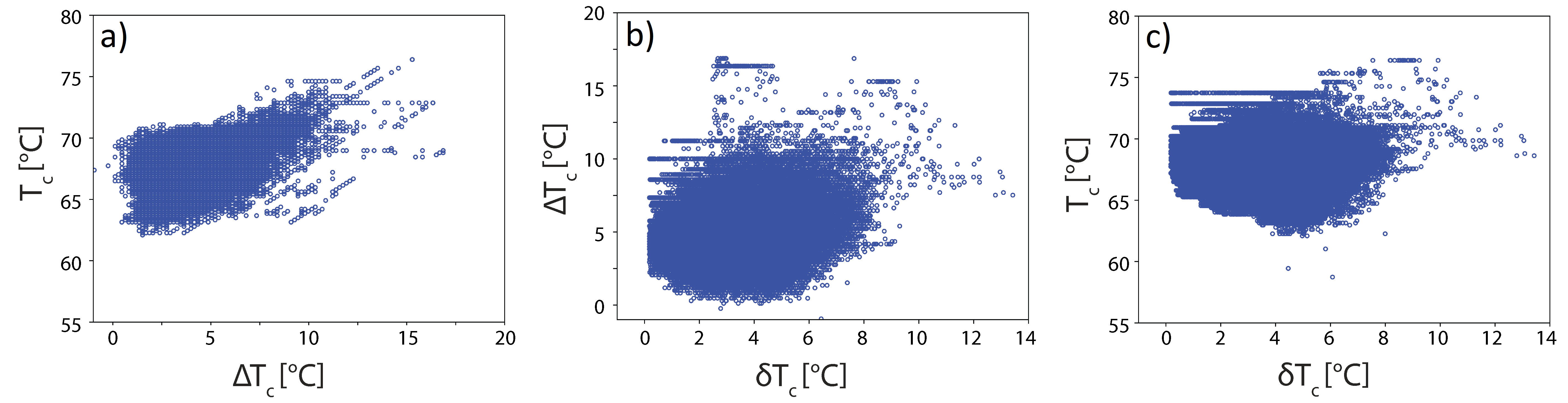}
  \caption{Correlations between T$_c$ (upon warming), $\Delta{\rm T}_c$ and $\delta{\rm T}_c$. Each of the 666,000 pixels (900x740) is represented. Only T$_c$ vs. $\Delta{\rm T}_c$ (panel (a) shows a slight diagonal trend meaning that pixels with low T$_c$ tend to have low $\Delta{\rm T}_c$ ({\em i.e.} close to zero) and {\em vice versa}.
\label{correlation}}
\end{figure*}

Having described the various key steps in the previous sections (including autofocusing, step motor/cross correlation aligning, single pixel scaling and thresholding, pair connectivity correlation length analysis, and time domain convolution) we now present the detailed spatially-resolved study of the IMT in VO$_2$ films using our new optical mapping method.

\subsection*{Maps}
{\bf Transition Temperature T$_c$ maps}: Fig.~\ref{fig:Maps} (c) reports the local critical temperature T$_c$ map in VO$_2$ sample B. These maps show a large spatial variation in T$_c$, with rich pattern formation
over tens of microns, similar to s-SNIM sub-micron measurements \cite{Qazilbash2007}, but acquired with a much faster procedure that allows for much finer time and temperature resolution. 
This large scale spatial variation, along
with detailed spatial knowledge of the location of these variations, 
can potentially be exploited to optimize memory elements by addressing specific 
regions of the sample.  

{\em Reproducibility of T$_c$ maps}:
Previous reports on avalanches in this material showed jumps in resistivity randomly appearing during the transition in macroscopic transport measurements \cite{Sharoni2008}. This suggested that the metal-insulator patterns could be appearing randomly during each temperature sweep. At first glance, this appears to be at odds with the optical data reported in this study, where we find that the metal and insulator patterns are highly repeatable globally (occurring at the same location and with the same shape) during successive temperature sweeps (see Fig \ref{fig:Repro}).  The repeatability suggests that 
the patterns are strongly influenced by an underlying random field present in the thin film or its substrate \cite{shuo-prl, lukasz-ML, MLpaper}. The observed stochasticity of resistance jumps in transport measurements \cite{Sharoni2008} could arise from small variations in the exact time at which avalanches are triggered.  In addition, small changes in optical maps can potentially create large changes in resistance, when tiny ``shorts'' connect pre-existing larger metallic clusters.

\begin{figure*}[ht!]
\centering
 \includegraphics[width=1.45\columnwidth]{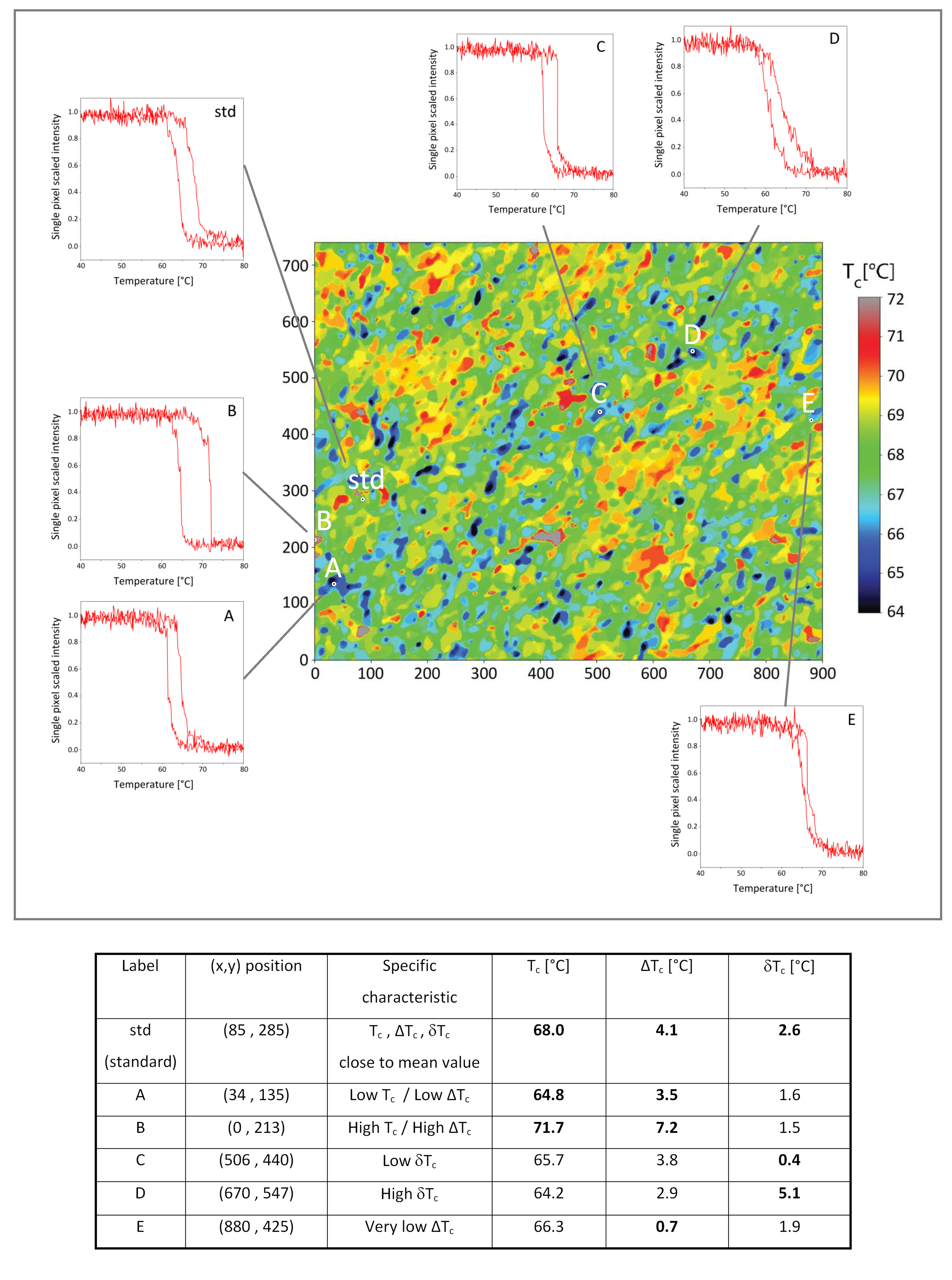}
  \caption{T$_c$ map with six pixels chosen to illustrate specific characteristics in the hysteresis loops. The table shows the numerical values of T$_c$, $\Delta$T$_c$ and $\delta$T$_c$ for each pixel. The numbers in bold highlight the unique characteristic of each pixel.
\label{Fig:PixelHunt}}
\end{figure*}

{\bf Transition Width $\delta$T$_c$ maps}: 
The transition width $\delta$T$_c$ of each pixel can be accessed by fitting single pixel scaled intensity time traces to a hyperbolic tangent: $-\frac{1}{2}(tanh$($\frac{{\rm T-T}_c}{\delta {\rm T}_c}$)-1). 
Because T$_c$ is known from our time trace analysis, there is only one fitting parameter.
The map of $\delta$T$_c$ distribution is shown in Fig.~\ref{fig:Maps} (e).
The average transition width of the pixels as measured in optics is $2.8 \pm 1.1$°C with extremes from 0°C to 8°C. Moreover, a small number of pixels show more than one step during a transition (see for example first pixel (305,300) in Fig. \ref{fig:TimeTraceSmall}).  
These cases could arise from an overlap between multiple metal or insulator domains affecting a single pixel.  
This could be due to information from surrounding pixels affecting the signal
at one pixel, since the pixel size is $\sim$10 times smaller than the resolution.
Or, it could arise from structures that are smaller than the pixel size.  Indeed, s-SNIM has clearly observed inhomogeneities on smaller length scales than the optical maps presented here \cite{Qazilbash2007,shuo-prl}.
Interestingly, 
the standard deviation of local T$_c$'s across the sample, $\sigma_{{\rm T}_c}$(1.2°C),
is smaller than the average transition width of pixels {$\delta$T$_c$}(2.8°C).
It remains an open question whether the self-similar metal-insulator domain patterns discussed in Ref.~\cite{shuo-prl} could be the source of this difference.  

{\bf Hysteresis Width $\Delta$T$_c$ maps}:
By subtracting T$_c$$^{up}$-T$_c$$^{down}$ (see the caption of Fig.\ref{fig:Maps} (b) for the definition) one can construct a hysteresis width $\Delta$T$_c$ map. The hysteresis width $\Delta$T$_c$ map is shown in Fig.~\ref{fig:Maps} (d) for sample B. The average width is found to be $4.3\pm 1.1$°C as seen in macroscopic transport measurements.  However, certain small regions have small $\Delta$T$_c$, in the range  [0°C - 1°C]  (small blue clusters in Fig.~\ref{fig:Maps} (d)). 
Probing these region with other local probes could shed light 
on whether %if/why 
this is an intrinsic property of these regions.
These hysteresis-free patches could be very useful in multiple switching applications such as optical electronic devices. Indeed it has been shown that the presence of a large hysteresis in VO$_2$ greatly complicates using it as an optical sensor \cite{Gurvitch2009}.

\subsection*{Correlations between maps}
With all of the maps above, one can check for correlations between these quantities. Fig.~\ref{correlation} plots T$_c$ vs. $\Delta{\rm T}_c$, $\Delta{\rm T}_c$ vs. $\delta{\rm T}_c$ and T$_c$ vs. $\delta{\rm T}_c$ for each pixel. 
A few horizontal and diagonal lines appear in these plots. The horizontal lines come from multiple pixels (spatially close by) switching at the same temperature (upon warming). The  diagonal lines come from multiple pixels (spatially close by) switching at the same temperature (upon cooling). 
Although this is typically what one would expect from avalanches, further analysis is needed to extract the full dynamics occurring.
In the three correlation maps, no trend is seen in the last two, but T$_c$ vs. $\Delta{\rm T}_c$ shows a slight positive correlation. This means that pixels with low T$_c$ tend to have low $\Delta{\rm T}_c$ ({\em i.e.} close to zero) and {\em vice versa}. The positive correlation in Fig.~\ref{correlation}(a) is not to be confused with the few diagonal lines present in this panel explained just above.

\subsection*{Hand picking specific hysteric properties}
The wide range of behaviors contained in the three maps presented in the section above (Fig.~\ref{fig:Maps} c, d and e), gives us the unprecedented opportunity to find individual pixels with desired properties. Fig.~\ref{Fig:PixelHunt} shows the T$_c$ map of the sample with six different types of pixels selected.
The pixel labeled “std” for standard has a rounded transition with values of T$_c$, $\Delta$T$_c$ and $\delta$T$_c$ which are close to the average values found in the distribution of these three quantities (see Fig.~\ref{Histograms} a, b and c).

Pixels A and B show the most common type of local characteristics found in the maps: when  T$_c$ is high,  $\Delta$T$_c$ is high; when T$_c$ is low, $\Delta$T$_c$ is low. This positive correlation is evident at a global level in Fig.~\ref{correlation} (a). 
However, on a local level, individual pixels can have a large deviation from the global average behavior.  
Indeed pixel E shows a possibility of finding $\Delta$T$_c$ very low (0.3°C) with a T$_c$ (66.3°C) low but closer to the mean value of the map.

Pixels C and D illustrate the case where the width $\delta$T$_c$ of the transition is very sharp (0.5°C) or very wide (5°C). Pixel C shows a representative sharp pixel,
where within the temperature steps of 0.17°C, the transition occurs in a sharp, avalanche mode. Further analysis to see where and how these avalanches occur will be pursued in future work.

Finally pixel E shows a case where $\Delta$T$_c$ is within the lower values [0°C-1°C]. As mentioned previously, small hysteresis could be useful in opto-electronic devices or neuromorphic devices. In the first case, small hysteresis avoids optical detectors getting stuck in subloops \cite{Gurvitch2009}; in the second case, small hysteresis allows lowering the voltage threshold needed for spiking \cite{Maffezzoni2015}.

{\em General remarks} on the pixel selection procedure: ($i$) as mentioned previously in the $\delta{\rm T}_c$ section above, some pixels in the map clearly present two steps during the IMT. 
These two-step pixels can potentially be detected in an automated way from their anomalously high error on the fit to the hyperbolic tangent function; ($ii$) the features put forward in these 6 pixels above are not unique to the 37$nm$ square pixel location. These features usually also hold for many pixels around the $xy$ coordinates reported.

\section{Conclusions} 

We have reported the first T$_c$ maps derived from
single pixel optical imaging on VO$_2$. Multiple new experimental steps were needed to align, focus and calibrate the raw grayscale images recorded. These experimental achievements allowed us to accurately track the spatial distribution of metal and insulator clusters. Binary black and white images, time traces, T$_c$ maps, $\Delta$T$_c$ maps, and $\delta$T$_c$ maps were plotted and discussed. The sample shows micron-sized  patterns that are found to be mostly reproducible through multiple temperature sweeps. The $\Delta{\rm T}_c$ hysteresis width map exhibits, on average, the same average hysteresis width of 4.3°C as macroscopic resistivity hysteresis, but exhibits strong variation on a local scale,
down to $\sim$[0°C-1°C] in certain small regions and as 
large as $\sim$ 8°C in other regions. These findings open an exciting opportunity to access local properties of VO$_2$ by, {\em e.g.}, contacting specific parts of the sample electrically in order to select unique parameter combinations for specific applications in electrical and optoelectronic devices. The observation of a positive correlation between T$_c$ value and hysteresis width could enable a new approach for tailoring the material’s response to external drives, in addition to providing a new perspective in studying open questions in the theory of hysteresis.

\section*{Acknowledgements}  

We thank M.~J.~Carlson for technical assistance with image stabilization, and acknowledge helpful conversations with K.~A.~Dahmen. S.B., F.S., and E.W.C. acknowledge support from NSF Grant No. DMR-2006192 and the Research Corporation for Science Advancement Cottrell SEED Award. S.B. acknowledges support from a Bilsland Dissertation Fellowship. E.W.C. acknowledges support from a Fulbright Fellowship, and thanks the Laboratoire de Physique et d'\'{E}tude des Mat\'{e}riaux (LPEM) at \'{E}cole Sup\'{e}rieure de Physique et de Chimie Industrielles de la Ville de Paris (ESPCI) for hospitality.
This research was supported in part through computational resources provided by Research Computing at Purdue, West Lafayette, Indiana {\cite{rcac-purdue}}.
The work at UCSD (PS, IKS) was supported by the Air Force Office of Scientific Research under award number FA9550-20-1-0242.
The work at ESPCI (M.A.B., L.A., and A.Z.) was supported by
Cofund AI4theSciences hosted by PSL University, through the European Union’s Horizon 2020 Research and Innovation Programme under the Marie Skłodowska-Curie Grant No. 945304.

\bibliographystyle{apsrev4-1}
\bibliography{Correlative_mapping_of_local_hysteresis_properties_in_VO2.bib}

\clearpage

%%%% Supporting Information section setting %%%%
\setcounter{subsection}{0}
\setcounter{equation}{0} % added by Yuxin
\makeatletter 
\renewcommand{\thesubsection}{S\arabic{subsection}}
\renewcommand{\theequation}{S\arabic{equation}}
\makeatother
%%%% Supporting Information figure setting %%%%

%\textbf{Supporting Information} %%%%
\section*{Supporting Information: Correlative mapping of local hysteresis properties in VO$_2$} 

%%%% Supporting Information figure setting %%%%
\setcounter{figure}{0}
\makeatletter 
\renewcommand{\thefigure}{S\arabic{figure}}
\makeatother
%%%% Supporting Information figure setting %%%%

\subsection{VO$_2$ Reflectivity}
\label{sxn:refelctivitySI}

\begin{figure}[hb!]
\centering
 \includegraphics[width=0.45\textwidth]{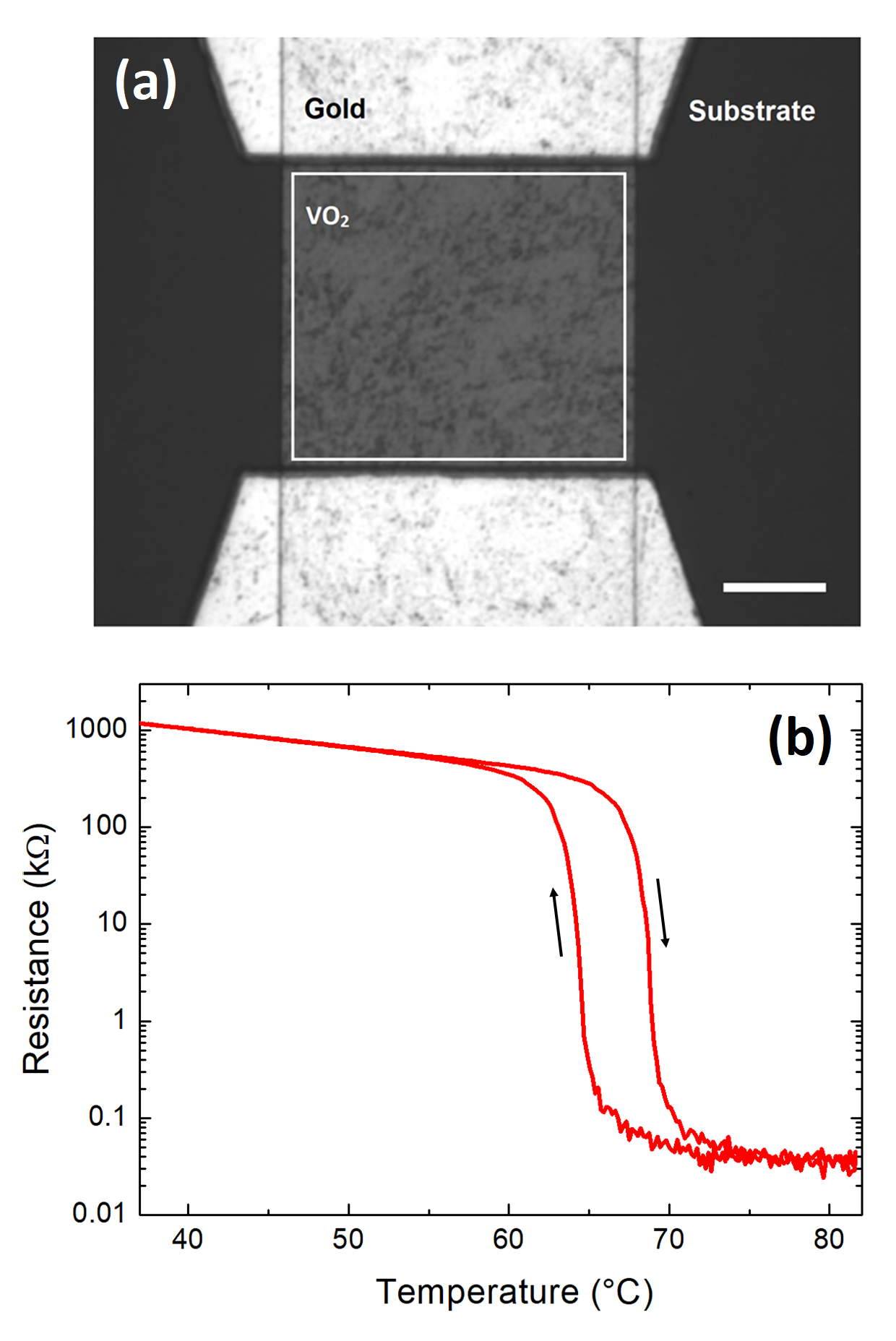}
  \caption{(a) 35$\mu m$ wide etched VO$_2$ sample B image with 30$\mu m$ separated gap gold leads. The white square represents the 33.6$\mu m$ x 27.6$\mu m$ region where T$_c$ maps (Fig.s~\ref{fig:Maps}). Scale bar is  10$\mu m$.(a) R(T) measurement of the IMT 
\label{fig:resistance}}
\end{figure}

\begin{figure*}[ht!]
\centering
 \includegraphics[width=0.55\textwidth]{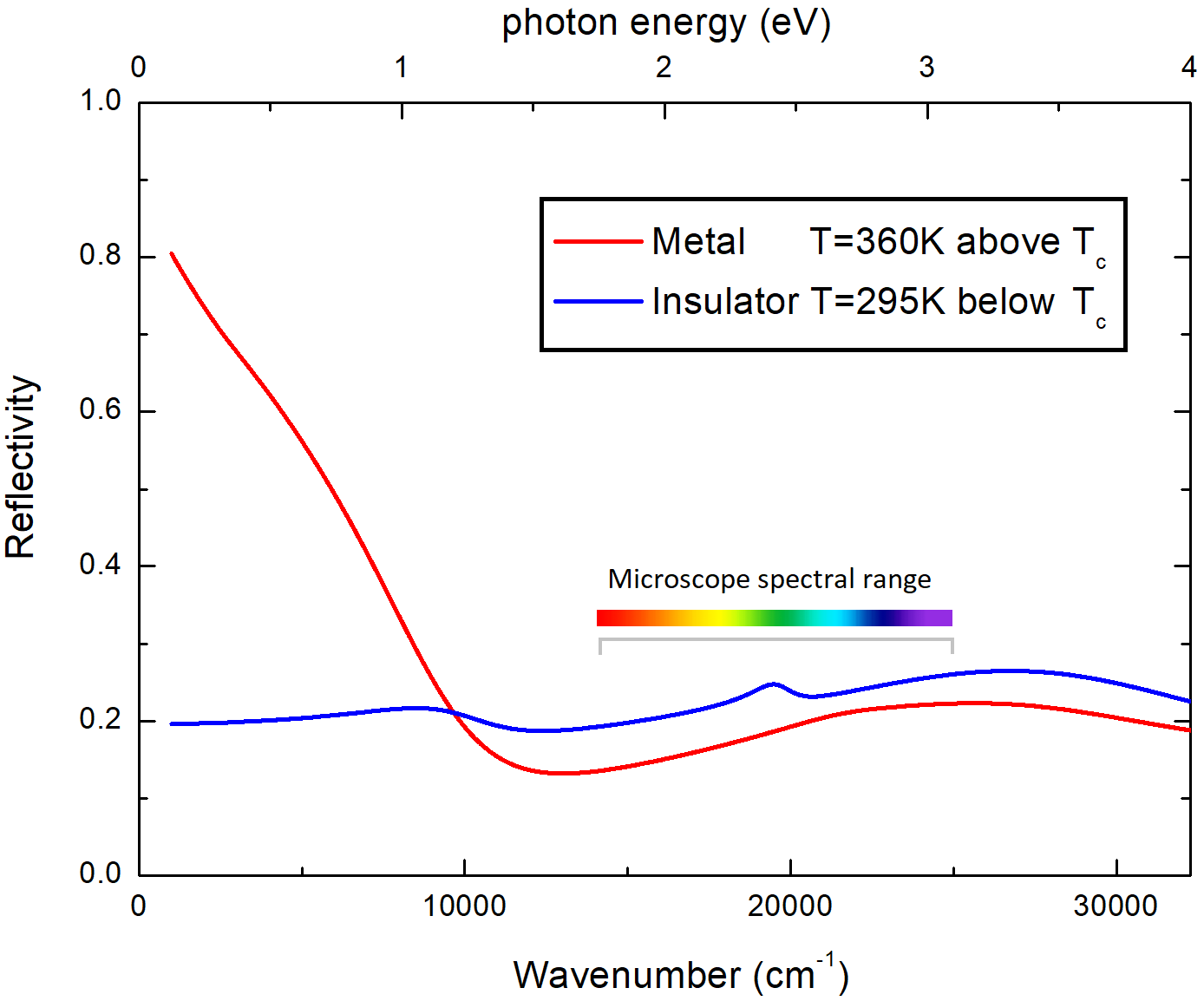}
  \caption{Simulated optical reflectivity of the insulating and metallic states in bulk VO$_2$. Optical functions were derived by fitting standard Drude-Lorentz functions to  ellipsometry measurements reporting the raw $\sigma_1$ response in a large spectral range at low and high temperatures \cite{Qazilbash_Optics2008}. This procedure \cite{Zimmers2005} allows  other optical functions to be deduced, such as reflectivity, transmission, absorption, or dielectric constant.  Reflectivities  in this figure are not reported below 1000$cm^{-1}$ as the fitting procedure was not precise enough in this low frequency/high $\sigma_1$ region. On the other hand, reflectivities in the visible region ($\sim$14000cm$^{-1}$ to $\sim$25000cm$^{-1}$) are in the middle of the spectral range and can be found with confidence.}
\label{fig:ReflectivityVO2}
\end{figure*}

\begin{figure*}[ht!]
\centering
 \includegraphics[width=0.55\textwidth]{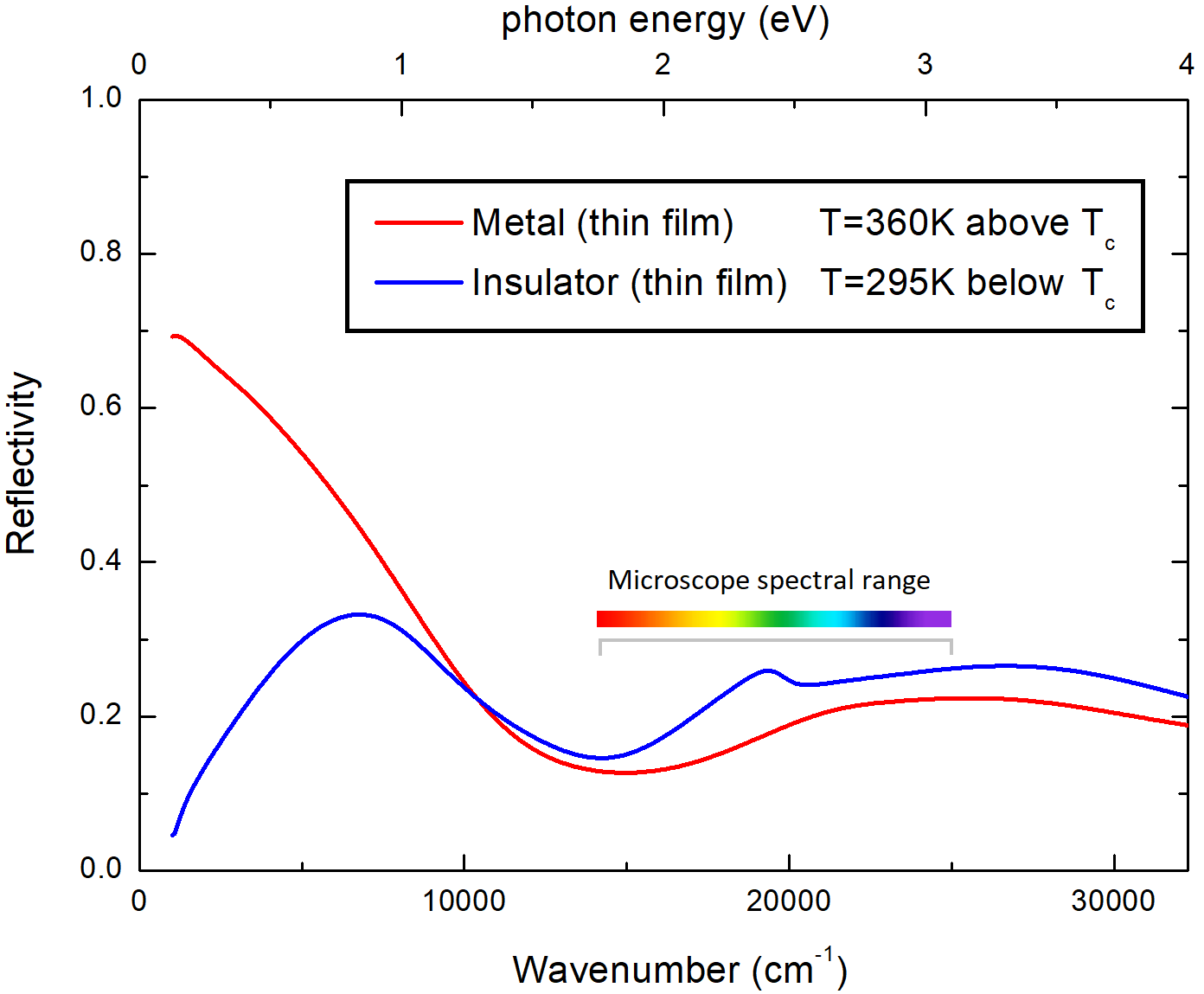}
  \caption{Simulated optical reflectivity of the insulating and metallic states of a 130$nm$ VO$_2$ thin film on an r-cut sapphire substrate. Optical functions were found as described in Fig.~\ref{fig:ReflectivityVO2}. 
  In contrast with the bulk reflectivity, a pronounced oscillation 
  can be seen  in the  blue insulating spectrum. This is due to  interference in the 130$nm$ thin film (for example, constructive thin film interference creates a peak at $\sim$6700cm$^{-1}$). Reflectivities are not reported below 1000$cm^{-1}$ as the fitting procedure was not precise enough in this low frequency/high $\sigma_1$ region. On the other hand, reflectivities in the visible region ($\sim$14000cm$^{-1}$ to $\sim$25000cm$^{-1}$)  are in the middle of the spectral range and can be found with confidence.}
\label{fig:ReflectivityVO2Film}
\end{figure*}

The fact that the metallic reflectivity of VO$_2$ is lower than that of the insulating phase in the visible range is counterintuitive. 
This is due to a subtle combination of a Drude response as well as intraband and interband transitions and thin film interferences in this material. 
The largest reported spectra in VO$_2$ was measured by ellipsometry \cite{Qazilbash_Optics2008}. Using the reported real part of the optical conductivity $\sigma_1$, we have calculated the reflectivity of the insulator and metallic states (see Fig. \ref{fig:ReflectivityVO2} and \ref{fig:ReflectivityVO2Film}). This clearly shows that, as one would expect in the infrared, the sample becomes highly reflective when metallic. Above the plasma frequency ($\sim$12000$cm^{-1}$), interband transitions and spectral weight conservation make the reflectivity curves cross, leading to the metallic state having a lower reflectivity than the insulating state in this range. The relative optical contrast in the visible range (27\%), is still more than sufficient in our setup to identify both states clearly (as seen in a raw image Fig.~\ref{fig:resistance} (a)).\\

\subsection{Key steps making this study possible}
\label{sxn:Key}
 The key step that have allowed us completing this study comes from the unique qualities of the VO$_2$ material :

- The IMT is above room temperature, which allows close optical microscopy (strong objective $\times$150 with a high numerical aperture 0.9 brought to 1$mm$ focus above the sample surface). This setup would be much harder to achieve if cryogenic cooling ({\em i.e.} a cryostat with a window between the sample and objective) was needed.

- Phase separation was observed by s-SNIM at sub-micron scales in this material \cite{Qazilbash2007,shuo-prl}. The fact that this phase separation is still found up to 30$\mu m$ makes these optical microscopy surface maps possible.

- In the visible range, a relative 27\% drop in the thin film reflectivity is found in the metallic state 
Measuring in the visible range gave us results with a 400$nm$ resolution. In the infrared, the contrast between metal and insulator is much larger, as expected, but only allows optical resolution up to the IR wavelength, {\em i.e.} 1-10$\mu m$.

\clearpage

\subsection{Image sensitivity drift correction}
\label{sxn:sensitivitySI}
Whereas the relative average intensity of VO$_2$ increases almost 30\% in changing from metal to insulator, the change in sapphire reflectance in this temperature range is negligible. We have used this fact to correct for any changes in incident light or CCD detector sensitivity throughout the experiment by dividing the average intensity in the VO$_2$ region by the intensity in the sapphire region of the sample.

Details: We assume that the input intensity is a function of time $I_0(t)$ but spatially uniform.
The reflected intensity from any region is $I_R(t,x,y) = I_0(t) \times R(t,x,y)$. Since the Sapphire's reflectance does not vary significantly over the range of temperature the sample went through, it is assumed to be a constant.
Let the spatially averaged sapphire reflectivity be $R^S$.
Then, the spatial average reflected intensity from the sapphire region is:
$ I_R^S(t) = I_0(t) \times R^S $
Any region of VO$_2$ has a reflected intensity:
$I_R^V(t,x,y) = I_0(t) \times R^V(t,x,y)$.
Therefore, the ratio of reflected intensities from Sapphire and VO$_2$ is independent of input intensity:
${I_R^V(t,x,y)}/{I_R^S(t)} = {R^V(t,x,y)}/{ R^S }$.
We will use $I_R^S(t)$ as a reference to correct $I_R^V(t)$ for any variation due to fluctuation of ambient light.
The quantity independent of input intensity:
$R^V(t,x,y) = { R^S }{I_R^V(t,x,y)}/{I_R^S(t)} $,
Hence, setting the reference input intensity $I_0(0)$, the corrected reflected intensity from VO$_2$ would be:
$$ \tilde{I}_R^V(t,x,y) = I_0(0)R^V(t,x,y) = \frac{I_R^V(t,x,y)}{I_R^S(t)/I_R^S(0)} $$

\subsection{Single pixel thresholded images: inflection point}
\label{sxn:inflectionSI} 
In the main text, we have set the threshold between metal and insulator domains at the midway point of the intensity, based on the pair connectivity correlation length criterion described in Sec.~\ref{sxn:thresholdSI}.
We also tested another method of setting the threshold based on
the inflection point of the single pixel time traces.
The green curves in panels (a-l) of Fig.~\ref{fig:mini_steps} show a smoothed derivative of the raw time traces, achieved by using a finite difference with a 11-point Gaussian convolution ($\sigma$=2.5) \cite{finite-difference}. The vertical dotted green line shows the extremum of this derivative, which locates the inflection point of the orange curves. Since the pixel switching curves (orange and blue traces) exhibit a relatively rapid change from metal to insulator, this inflection point at which the pixel brightness is changing most rapidly is the most natural place to assign a change from insulator to metal and {\em vice versa}. Because we have used a stencil with even number of 10, the inflection point happens between frames, and allows us to clearly identify frames which precede the inflection point (which are metallic) from frames which come after the inflection point (which are insulating). 
Notice that the frame number at which the solid orange curves cross the dotted orange lines coincides with the inflection point for each pixel. This means that both methods are equivalent for determining the frame number at which a pixel switches from metal to insulator or {\em vice versa}. \\

\begin{figure*}[hb!]
\centering
 \includegraphics[width=0.7\textwidth, clip]
{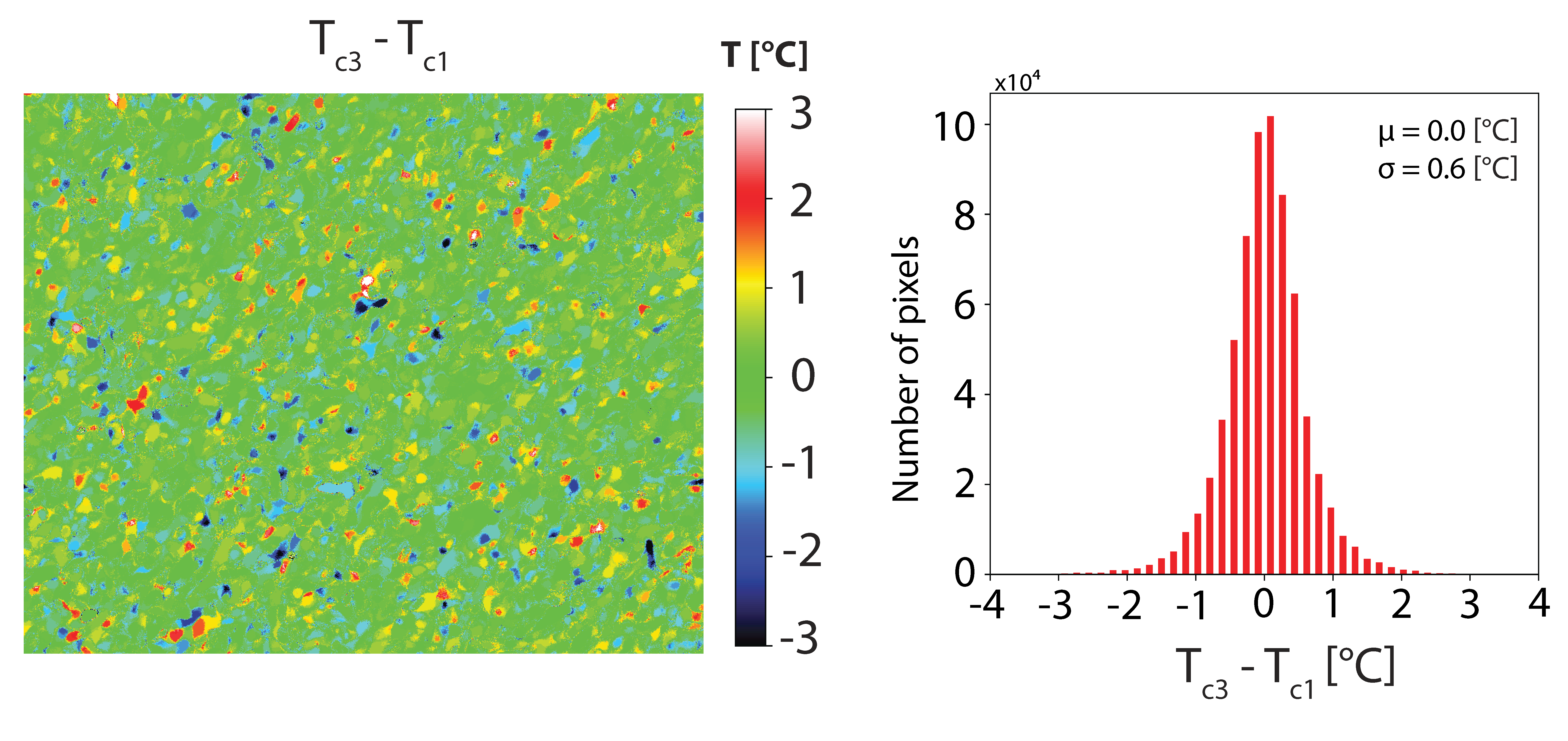}
  \caption{Map and histogram of the difference between T$_{c3}$ and T$_{c1}$. Although these T$_c$ maps are the most separated, time wise, in this study, they remain similar (mean and $\sigma$) to T$_{c2}$-T$_{c1}$ and T$_{c3}$-T$_{c2}$ presented in Fig.\ref{fig:Repro}.
\label{Fig:fullTcDiff}}
\end{figure*}

\begin{figure*}[hb!]
\centering
\includegraphics[width=0.9\textwidth, clip]
{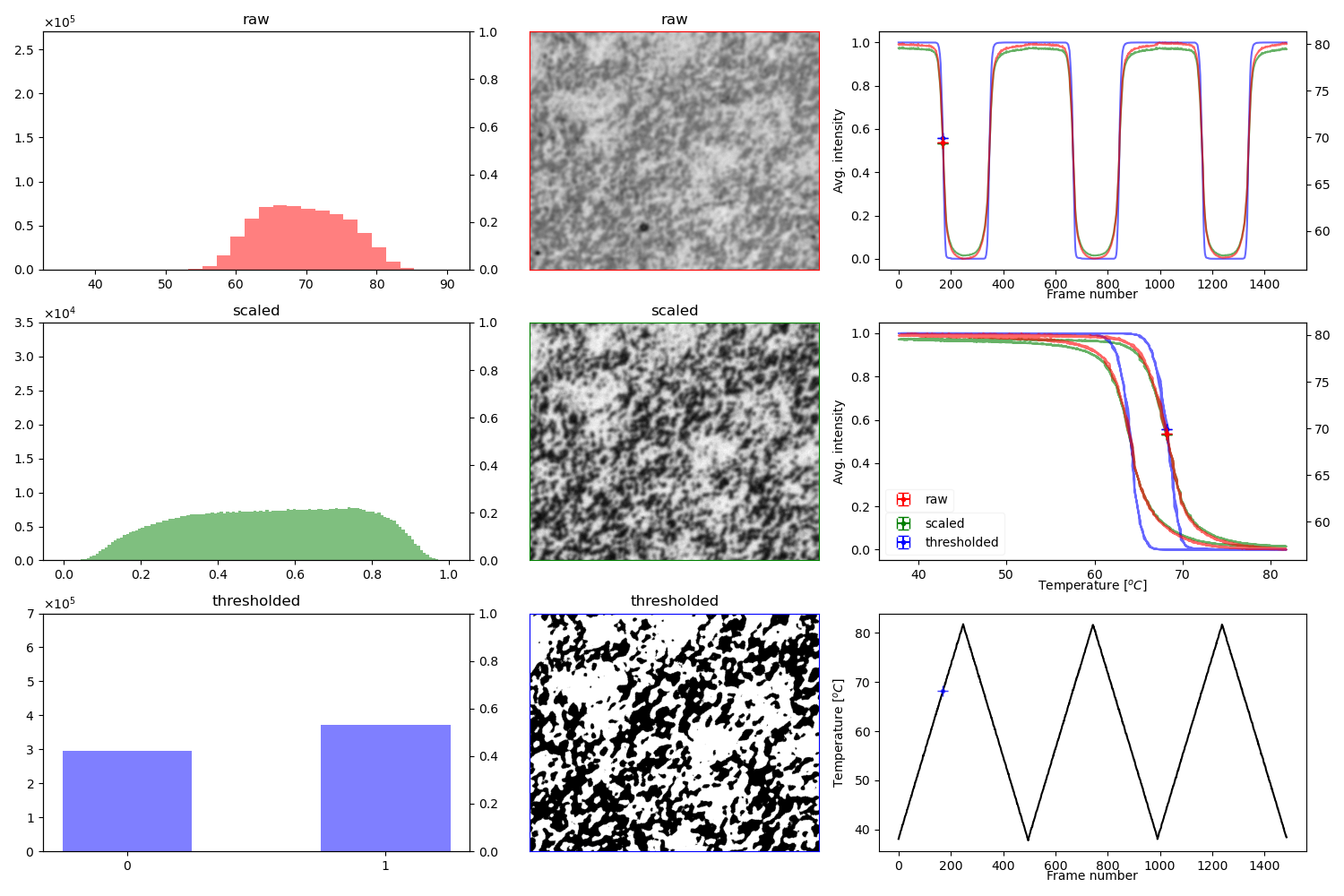}
  \caption{\textbf{Online} \textbf{movie}\textsuperscript{\cite{YouTubeMovie}} screenshot of the $\sim$1500 in focus consecutive spatial maps  of a 33.6$\mu m$ x 27.6$\mu m$ VO$_2$ surface.
  Central panels: raw, scaled and thresholded surface image (sample B) using ``Single pixel scaled image'' and ``Single pixel intensity time trace and threshold'' methods.
  Left panels: corresponding histogram changes during temperature ramps.
  Top right panel: average sample intensity (raw, scaled, thresholded) vs. frame number. Middle right panel: average sample intensity (raw, scaled, thresholded) vs. sample temperature. Bottom right panel: Temperature protocol - 3 major temperature loop spanning the entire IMT (36$^o$C - 82$^o$C),
\label{3IMTMovie}}
\end{figure*}

\begin{figure*}[hb!]
\centering
\includegraphics[width=\textwidth, clip]{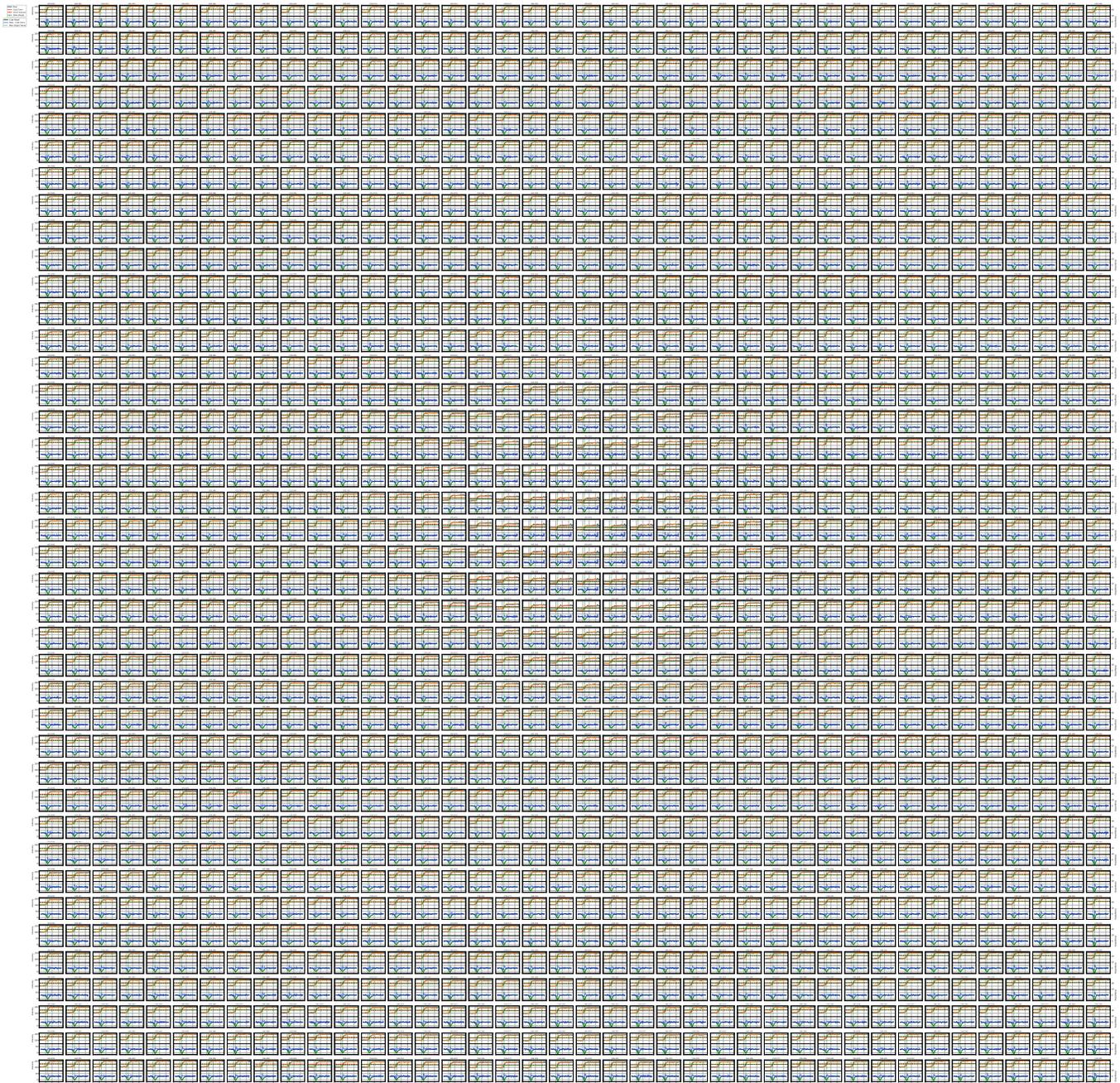}
  \caption{ML3 time trace of sample A in a patch of 40$\times$40 pixels in the middle of the sample.   
Each pixel coordinates are indicated above the time trace. Description of the four curves in each mini panel is the same as the main text Figure~\ref{fig:mini_steps}.
\label{fig:TimeTraceSmall}}
\end{figure*}

\end{document}